\documentclass[11pt]{article}
\pdfoutput=1

\usepackage{jheppub}
\usepackage{url,comment}
\usepackage{times}
\usepackage{latexsym}
\usepackage{graphicx, graphics, hyperref, amsmath, amssymb, slashed, xcolor,bm,amsthm, array}
 \usepackage{subfigure}
 \usepackage{listings}
 \usepackage{trimclip}

\usepackage{rotating}
\usepackage{afterpage}
\newcommand{\nc}{\newcommand}

\nc{\beq}{\begin{equation}}
\nc{\eeq}{\end{equation}}
\nc{\barray}{\begin{eqnarray}}
\nc{\earray}{\end{eqnarray}}
\nc{\barrayn}{\begin{eqnarray*}}
\nc{\earrayn}{\end{eqnarray*}}
\nc{\bcenter}{\begin{center}}
\nc{\ecenter}{\end{center}}
\nc{\mc}{\mathcal}
\nc{\er}[1]{(\ref{eq:#1})}
\nc{\onehalf}{\frac{1}{2}} 
\nc{\partialbar}{\bar{\partial}}
\nc{\psit}{\widetilde{\psi}}
\nc{\Tr}{\mbox{Tr}}
\nc{\hc}{\mbox{H.c.}}
\nc{\ev}{\;\mathrm{eV}}
\nc{\mev}{\;\mathrm{MeV}}
\nc{\gev}{\;\mathrm{GeV}}
\nc{\tev}{\;\mathrm{TeV}}

\def\chii0{\chi_i^0}
\def\chij0{\chi_j^0}

\newcommand{\gsim}{\lower.7ex\hbox{$\;\stackrel{\textstyle>}{\sim}\;$}}
\newcommand{\lsim}{\lower.7ex\hbox{$\;\stackrel{\textstyle<}{\sim}\;$}}
\nc{\ttbar}{t\bar t}

\newcommand{\cref}[1]{Chapter~\ref{c.#1}}

\def\tw{\omega_d}

\graphicspath{{plots/}}

\title{Detector-size Upper Bounds on Dark Hadron Lifetime from Cosmology}

\author{Lingfeng Li$^{1}$ and Yuhsin Tsai$^2$}
\affiliation{$^{1}$Jockey Club Institute for Advanced Study, Hong Kong
University of Science and Technology, Hong Kong\\
$^{2}$Maryland Center for Fundamental Physics, Department of Physics, University of Maryland, College Park, MD 20742, USA
}

\abstract{
We show that in a confining hidden valley model where the lightest hidden particles are dark hadrons that have mass splittings larger than $\mathcal{O}(0.1)$ GeV, if the lightest dark hadron is either stable or decays into Standard Model (SM) hadrons/charged leptons during the big-bang nucleosynthesis (BBN), at least one of the heavier dark hadrons needs to decay into SM particles within $\mathcal{O}(10)$ nanosec. Once being produced at collider experiments, this heavier dark hadron is likely to decay within $\mathcal{O}(1)$ meter distance, which strengthens the motivation of searching for long-lived particles with sub-meter scale decay lengths at colliders. To illustrate the idea, we study the lifetime constraint in scenarios where the lightest dark particle is a pseudo-scalar meson, and dark hadrons couple to SM particles either through kinetic mixing between the SM and dark photons or by mixing between the SM and dark Higgs. We study the annihilation and decay of dark hadrons in a thermal bath and calculate upper bounds on the lightest vector meson (scalar hadron) lifetime in the kinetic mixing (Higgs portal) scenario. We discuss the application of these lifetime constraints in long-lived particle searches that use the LHCb VELO or the ATLAS/CMS inner detectors.
}

\begin{document}

\begin{flushright}
\small{.}
\end{flushright}

\maketitle

\section{Introduction}
A Confining Hidden Valley (CHV) containing dark hadrons which weakly couple to SM particles appears in many beyond the SM scenarios. Such CHV models have been used to solve the Higgs hierarchy problem~\cite{Chacko:2005pe,Burdman:2006tz,Craig:2015pha,Arkani-Hamed:2016rle,Cheng:2018gvu,Cohen:2018mgv}, strong CP problem~\cite{Hook:2014cda,Agrawal:2017ksf}, and address astrophysical and cosmological anomalies~\cite{Freytsis:2014sua,Hochberg:2014dra,Freytsis:2016dgf,Prilepina:2016rlq,Chacko:2018vss}. Neutral hadrons in a CHV sector which are not stabilized by a symmetry can slowly decay into SM particles. The resulting long-lived particle (LLP) signatures have motivated active research programs aiming to improve searches in the existing detectors~\cite{Bai:2013xga,Schwaller:2015gea,Curtin:2015fna,Csaki:2015fba,Chacko:2015fbc,Pierce:2017taw,Li:2017xyf,Kribs:2018ilo} and to propose new experiments~\cite{Curtin:2018mvb,Feng:2017uoz,Alekhin:2015byh,Gligorov:2017nwh,Liu:2018wte}. Existing searches from the LHCb, ATLAS, and CMS collaborations have already set useful constrains on dark hadron production, see e.g.~\cite{Lee:2018pag} for a review of the previous searches, and~\cite{Aaij:2017mic,Aaboud:2018aqj,Sirunyan:2018vlw,Sirunyan:2018njd} for some recent results.

The efficiency of LLP search depends on the size of the particle detector relative to the LLP decay length. Due to the small coupling between the hidden and SM sectors, the possible range of dark hadron decay lengths can range from $\mathcal{O}(10^{-6}{\rm}-10^8)$ meters. Such and even slower decays in the early universe can cause reheating and violate bounds from the Big-Bang Nucleosynthesis (BBN) and Cosmic Microwave Background (CMB) physics~\cite{Kawasaki:2017bqm,Poulin:2016anj}. However, the $10^8$ meter decay length is still way bigger than the particle detector in any collider experiment. If most CHV scenarios prefer such long hadron lifetimes, the efficiency of CHV searches is greatly suppressed by the tiny decay probability inside the detector. Larger detector such as MATHUSLA~\cite{Chou:2016lxi,Alpigiani:2018fgd} which can probe LLPs with decay lengths close to the BBN time scale has been proposed. Nevertheless, smaller existing detectors, such as the LHCb VELO (Vertex Locator) and the inner detectors of the ATLAS/CMS experiments have provided excellent reconstruction of displaced decays with $\mathcal{O}(10)$ cm decay lengths. It is therefore important to identify the type of CHV scenarios that satisfy the cosmological constraints and have well-motivated dark hadron decay lengths within these detectors.

\begin{figure}
\begin{center}
\includegraphics[width=7cm]{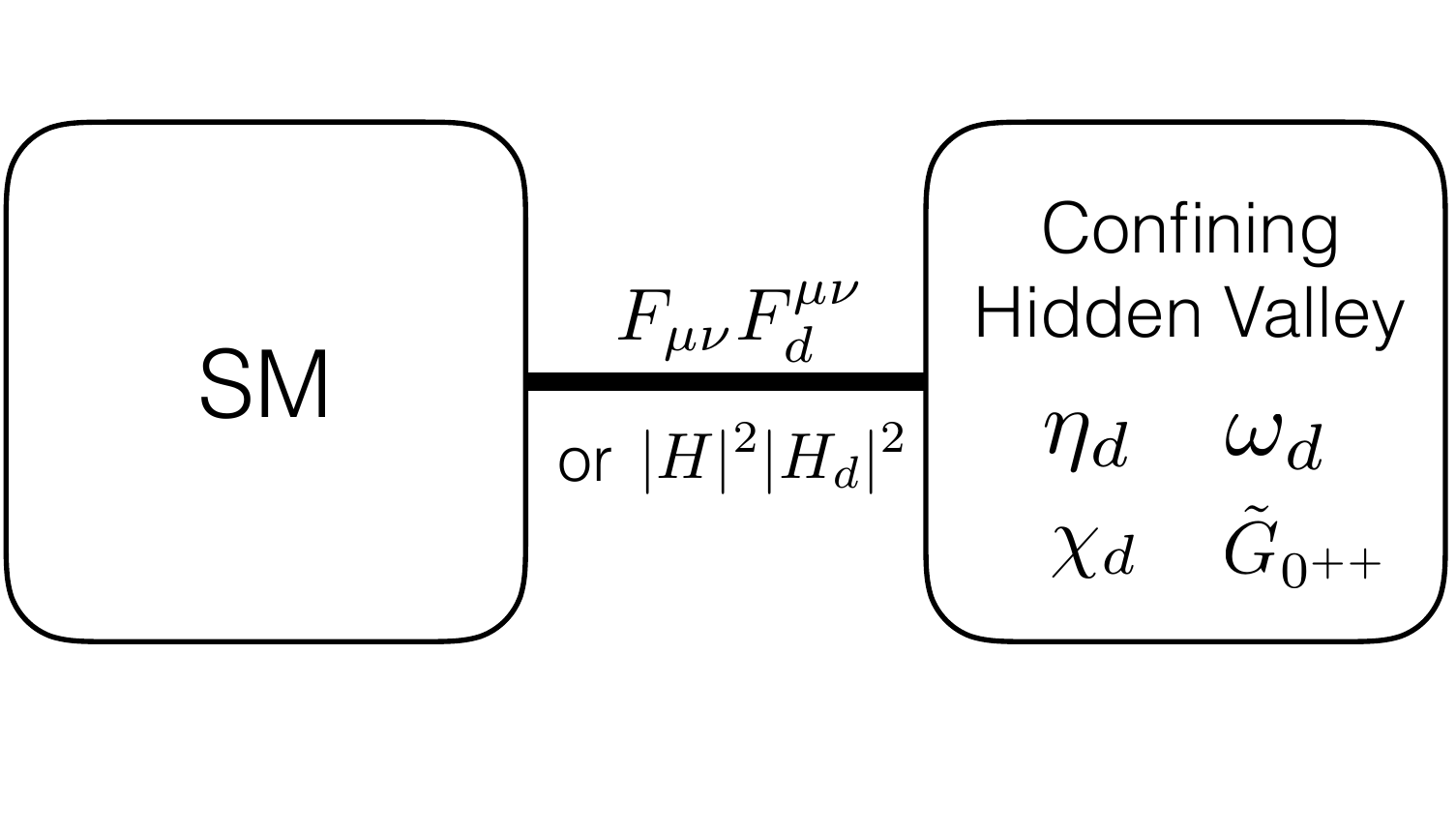}\qquad\quad\quad\includegraphics[width=7cm]{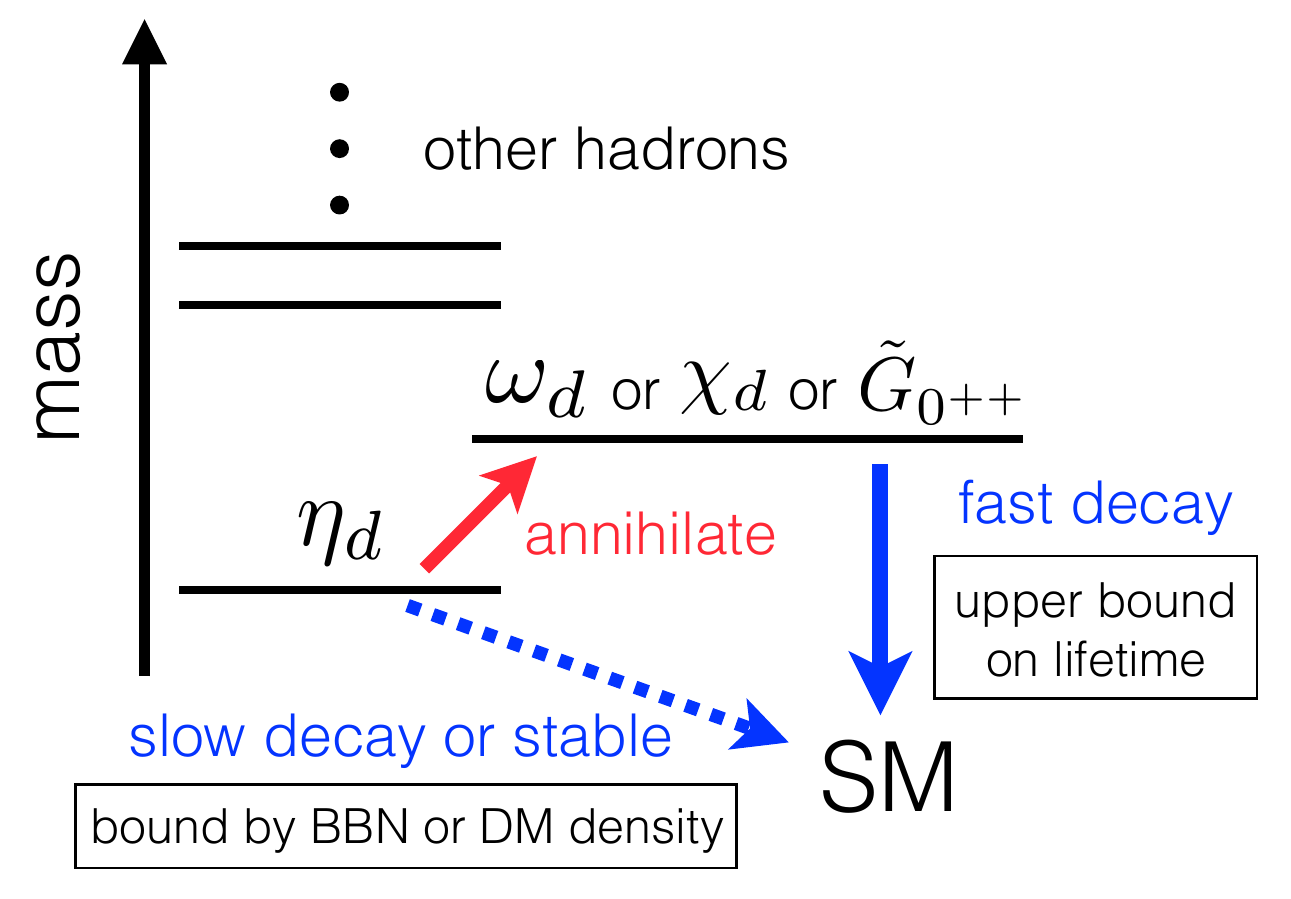}
\end{center}
\caption{\emph{Left:} the CHV scenarios studied in this work. The hidden sector couples to the SM sector either through the photon mixing or the Higgs portal coupling. The pseudo-scalar meson $\eta_d$ is the lightest particle in the dark confining phase. The lightest vector meson $\omega_d$, the lightest scalar meson $\chi_d$, and the dark glueball $\widetilde{G}_{0^{++}}$ play important roles in determining the dark hadron relic abundance. \emph{Right:} the process that reduces the $\eta_d$ energy. $\eta_d$ annihilates into heavier hadrons when the hidden sector temperature is comparable to the hadron mass splitting, and $\omega_d$ ($\chi_d$ or $\widetilde{G}_{0^{++}}$) decay quickly into SM particles through the photon (Higgs) portal coupling comparing to the Hubble time scale while the $\eta_d$ conversion is allowed. The remaining $\eta_d$ either decays into SM during the BBN/CMB time (photon mixing scenario), or becomes stable DM (Higgs portal scenario). The cosmological constraints on the $\eta_d$ abundance set upper bounds on the lifetime of $\omega_d$, $\chi_d$, or $\widetilde{G}_{0^{++}}$.}\label{fig:cartoon}
\end{figure}

In this work, we focus on dark sectors that have $\mathcal{O}(1{\rm -}10)$ GeV dark hadrons made of rather heavy fermionic dark quarks, and the dark confinement scale ($\Lambda_d$) is lower than the dark quark mass but higher than $15\%$ of the light dark meson mass. The lightest dark meson is a pseudo-scalar bound state ($\eta_d$), and the dark vector meson ($\omega_d$) and scalar meson ($\chi_d$) are heavier mesons in the triplet hyperfine and $p$-wave states~\cite{Nussinov:1999sx}. From the lattice calculation, the lightest glueball $\widetilde{G}_{0^{++}}$ has mass $\approx 7\Lambda_d$~\cite{Chen:2005mg}. The choice of dark confinement scale makes the lightest glueball heavier than $\eta_d$, so $\eta_d$'s do not annihilate into glueballs. We consider two types of CHV scenarios in which dark particles only couple to the SM sector either through a heavy dark photon or Higgs portal coupling (Fig.~\ref{fig:cartoon} left). As we show, in these scenarios, either $\omega_d$, $\chi_d$, or $\widetilde{G}_{0^{++}}$ has to have a proper decay length $\lsim10$ meter in order to satisfy the BBN and dark matter (DM) density constraints. We apply this lifetime constraint to collider searches of the exotic $Z$ or Higgs boson decay into LLPs and show that the LHCb VELO and the inner detector of ATLAS/CMS have good volume coverage of the displaced dark hadron decays. Moreover, the same cosmological constraints set lower bounds on the SM-dark sector coupling, which is complementary to collider searches.


The main motivation for the cosmological lifetime bound is the following. Since $\eta_d$ is the lightest hadron with thermal  abundance, it needs to decay into SM particles before $\sim 1$ sec in order to avoid disturbing the BBN process. However, since the pseudo-scalar meson either has $\gg 1$ sec lifetime through the decay via off-shell dark photons, or remains stable in the Higgs portal scenario, its relic density prior to $1$ sec needs to be much lower than the thermal abundance. Without invoking extra decay channels, the $\eta_d$ abundance can only be suppressed by first annihilating $\eta_d$ into heavier mesons ($\omega_d,\,\chi_d,\,\widetilde{G}_{0^{++}}$). The latter mesons then decay sufficiently quickly via the single photon or Higgs mediation to suppress the $\eta_d$ abundance before causing cosmological problems\footnote{Since we focus on dark hadrons above GeV scale, the $3\to 2$ annihilation process is not efficient enough to suppress $\eta_d$ abundance as in the strongly interacting massive particle (SIMP) scenario~\cite{Hochberg:2014dra}.}  (Fig.~\ref{fig:cartoon} right).

The conversion of $\eta_d$ into heavier mesons can happen only as long as the hidden sector temperature is larger than or comparable with the dark hadron mass splitting of $\mathcal{O}({\rm GeV})$ so as to allow fast $\eta_d\eta_d\to 2X$ ($X=\omega_d, \chi_d,\widetilde{G}_0$) conversion. Specifically, 
the conversion should last approximately the Hubble time scale $\sim 10^{-6}$ sec, and the decay of the $X$-hadron should be faster than that. Furthermore, the BBN constraints require that the co-moving number of the surviving $\eta_d$ before it decays into SM hadrons/leptons satisfy $Y_{\eta_d}\lsim10^{-9}$~\cite{Kawasaki:2017bqm}. The rate of decay of the slightly heavier dark mesons $\Gamma_{X}$ should therefore satisfy $Y_{\eta_d}\sim Y_{X}\sim\exp[-\Gamma_{X}/ H_{{\rm GeV}}]\lsim10^{-9}$, which is $c\tau_X\lsim10$ meter.

While the relation between $Y_{\eta_d}$ and $Y_{X}$ is more complicated than what we assume in this simple estimate, the main idea and results hold in the careful study below. The idea of setting a detector size upper bound on the LLP lifetime has been discussed in~\cite{Cheng:2015buv} to limit the parameter space for the twin upsilon search in the Fraternal Twin Higgs model~\cite{Craig:2015pha}. A similar thermal history of dark hadrons is also studied in~\cite{Beauchesne:2018myj,Berlin:2018tvf}. In this work, we perform more detailed study of the CHV models with more general assumptions of the dark hadron interaction and focus on the application of lifetime constraints to the LLP searches\footnote{See also~\cite{Cui:2012jh,Cui:2014twa} and \cite{Berlin:2018jbm} for the detector-size lifetime constraints on LLPs in the WIMP baryogenesis and inelastic DM scenarios.}.

The outline of this paper is as follows. In Sec.~\ref{sec:2} we introduce the Boltzmann equations that describe the evolution of dark hadron abundance. We provide analytical estimate of the hadron abundance, and the result works for a more general setup of CHV scenarios  described above. In Sec.~\ref{sec:3}, we apply the relic abundance study to the photon mixing and Higgs portal scenarios. We show constraints on hidden hadron lifetimes in both scenarios. In Sec.~\ref{sec:search}, we compare the lifetime time constraint to the size of LHCb VELO and ATLAS/CMS trackers, assuming dark hadrons are produced at these colliders through the exotic $Z$ or Higgs decay. Our conclusions are in Sec.~\ref{sec:conclusion}.

\section{Evolution of the hidden hadron density}\label{sec:2}

We first considering a CHV scenario where the lightest dark hadron $\phi_l$ and a heavier hadron $\phi_h$ have their comoving numbers evolve through the Boltzmann equations
\begin{eqnarray} 
\frac{d Y_{h}}{dx}&=&\frac{-1}{3H(x)}\frac{ds}{dx}\bigg[\left\langle\sigma_{+2h} v\right\rangle Y_{l}^2
-\left\langle\sigma_{-2h} v\right\rangle Y_{h}^2 -\left\langle\sigma_{-h} v\right\rangle Y_{h}Y_{l} +\left\langle\sigma_{+h} v\right\rangle Y_{l}^2\label{eqn:boltz_eta_1}
\\
&\quad&-\frac{\langle\Gamma_{\phi_h\to\text{SM}}\rangle_{\hat{T}}}{s}Y_{h} + \frac{\langle\Gamma_{\phi_h\to\text{SM}}\rangle_T}{s} Y_{h}^{\text{eq}}(T)\bigg],
\nonumber
\\
\frac{d Y_{l}}{dx}&=&\frac{-1}{3H(x)}\frac{ds}{dx}\bigg[\left\langle\sigma_{-2h} v\right\rangle Y_{h}^2 -\left\langle\sigma_{+2h} v\right\rangle Y_{l}^2+\left\langle\sigma_{-h} v\right\rangle Y_{h}Y_{l} -\left\langle\sigma_{+h} v\right\rangle Y_{l}^2
\bigg].\label{eqn:boltz_eta_2}
\end{eqnarray}
Here $Y_{l,h}=n_{l,h}/s$ is the comoving number of dark hadrons, and $s$ is the entropy density with its value determined by the SM temperature $T$. We define a dimensionless temperature variable $x\equiv m_h/T$ and assume the heavier hadron $\phi_h$ has a thermal averaged decay rate \cite{Bardakci:2009wp} into SM particles
\begin{equation}
\langle\Gamma_{\phi_h\to\text{SM}}\rangle|_{\hat{T},\,T}=\Gamma_{\phi_h\to\text{SM}} \bigg\langle\frac{m_h}{E_h}\bigg\rangle_{\hat{T},\,T} 
= \Gamma_{\phi_h\to\text{SM}} \frac{K_1(x)}{K_2(x)}\Big{|}_{x=\frac{m_h}{\hat{T}},\,\frac{m_h}{T}}\quad,
\end{equation} 
where $\hat{T}$ is the dark hadron temperature, and the thermally averaged dilation factor is given by
the ratio of the modified Bessel functions $K_{1,2}$. Since we consider scenarios in which the time scale of the $\phi_l$ freeze-out is much shorter than the $\phi_l$ lifetime, we neglect the $\phi_l$ decay in the Boltzmann equations.

In the dark sector, $\phi_{l,h}$ mesons are in chemical equilibrium due to the strong dark QCD interactions. The different average conversion cross sections, $\left\langle\sigma_{-2h} v\right\rangle$ for $\phi_h\phi_h\to\phi_l\phi_l$, $\left\langle\sigma_{+2h} v\right\rangle$ for $\phi_l\phi_l\to\phi_h\phi_h$, $\left\langle\sigma_{-h} v\right\rangle$ for $\phi_h\phi_l\to\phi_l\phi_l$, and $\left\langle\sigma_{+h} v\right\rangle$ for $\phi_l\phi_l\to\phi_l\phi_h$, relate to each other in thermal bath\footnote{The exponential suppression comes from a similar origin as in the forbidden DM~\cite{DAgnolo:2015ujb} model.} following ($\Delta m\equiv m_h-m_l$)
\begin{equation}\label{eq:Boltzmann}
\left\langle\sigma_{+2h} v\right\rangle = \frac{(Y_{h}^{\text{eq}})^2}{(Y_{l}^{\text{eq}})^2}\left\langle\sigma_{-2h} v\right\rangle \simeq \left\langle\sigma_{-2h} v\right\rangle e^{-\frac{2\Delta m}{\hat{T}}},\qquad
\left\langle\sigma_{+h} v\right\rangle = \frac{Y_{l}^{\text{eq}}\,Y_{h}^{\text{eq}}}{(Y_{l}^{\text{eq}})^2}\left\langle\sigma_{-h} v\right\rangle \simeq \left\langle\sigma_{-h} v\right\rangle e^{-\frac{\Delta m}{\hat{T}}}.
\end{equation}
We numerically check the inclusion of other conversion processes such as $\phi_h\phi_l\to\phi_h\phi_h$ and find them giving negligible corrections to the $Y_l$ evolution due to the smaller $\phi_h$ density in the initial state. Dark sector temperature $\hat{T}$ follows SM temperature $T$ mainly via the decay and inverse-decay of $\phi_h\leftrightarrow$ SM process until it decouples, and we calculate $\hat{T}$ by solving~\cite{Ma:1995ey,Buen-Abad:2015ova}
\begin{equation}
\frac{d\,\hat{T}}{dt}=-2 aH\hat{T} +\frac{2}{3} \frac{dE}{dt}\frac{\hat{T}-T}{\hat{T}},
\label{eq:Tdec2}
\end{equation}
where $t$ is the physical time. The temperature evolution basically follows 
\begin{equation}
\hat{T} = \begin{cases} T\, & (T\geq T_{dec})\, \\ \frac{T^2}{T_{dec}}\, & (T<T_{dec})\, \end{cases}
\end{equation}
for $T_{dec}\ll m_{h,l}$, and the $T_{dec}$ at the kinetic decoupling between the SM and dark sectors can be estimated 
by solving
\begin{equation}\label{eq:Tdec}
T^{-1}\,\frac{dE}{dt}\Big|_{T=T_{dec}}=H(T_{dec}).
\end{equation}
Since the $\phi_l$ abundance is determined by the decoupling of the $+(2)h$ conversion process, the surviving $Y_l$ is very sensitive to the $\hat{T}$ evolution. $Y_l$ freezes out quickly after $T< T_{dec}$. 

\begin{figure}
\centering
\includegraphics[height=6cm]{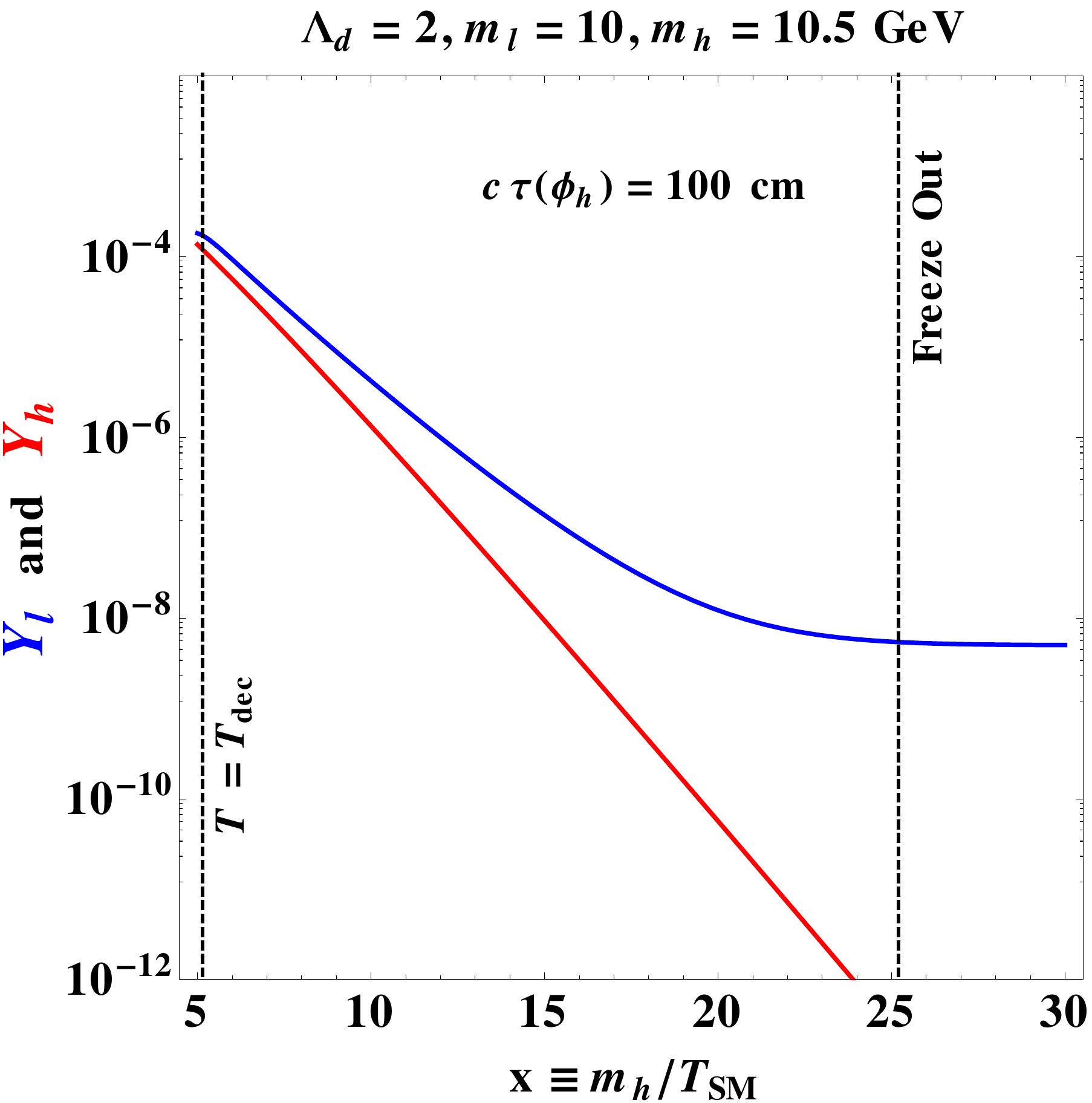}\qquad
\includegraphics[height=6cm]{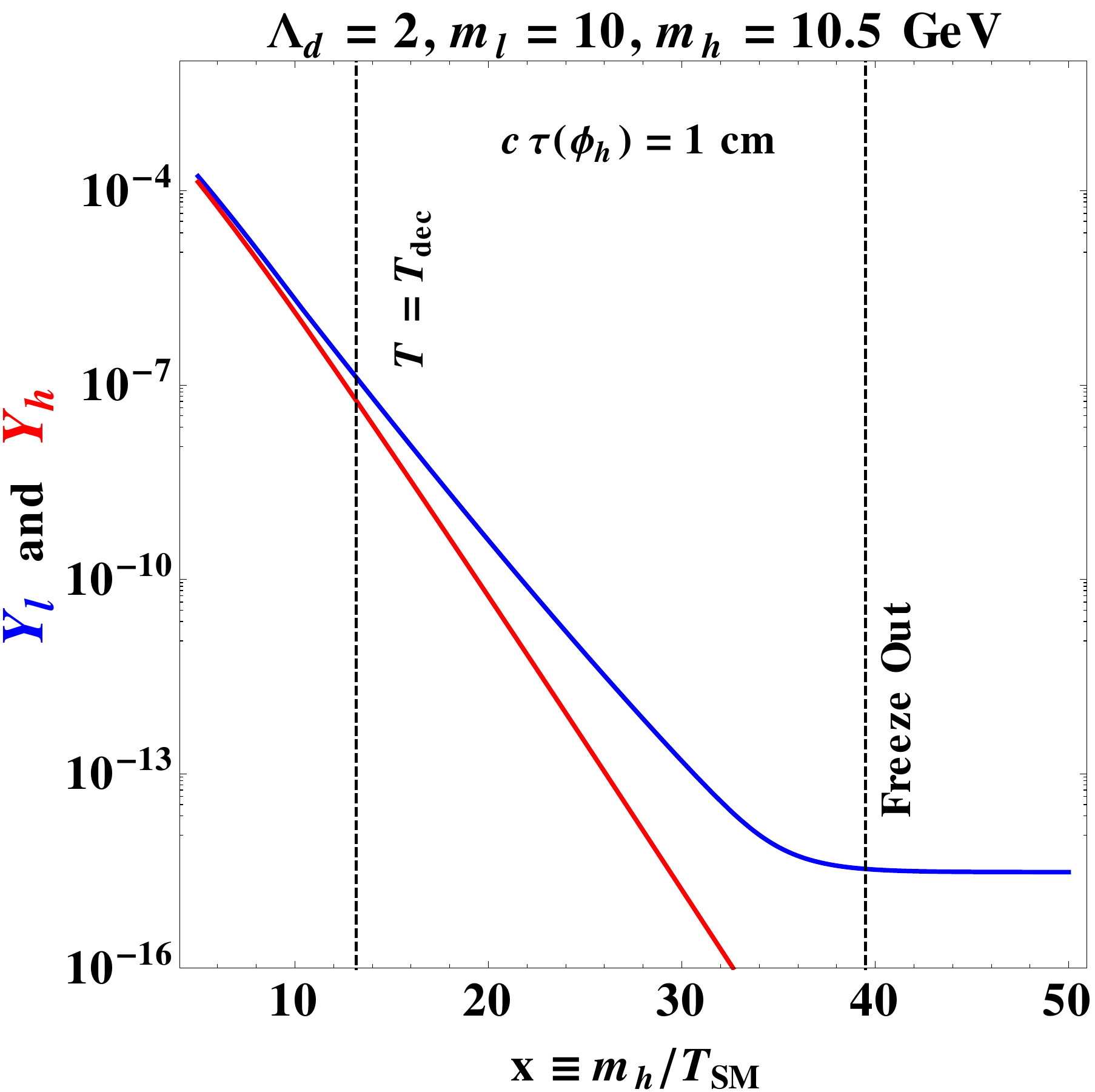}
\\\vspace{1em}
\includegraphics[height=6cm]{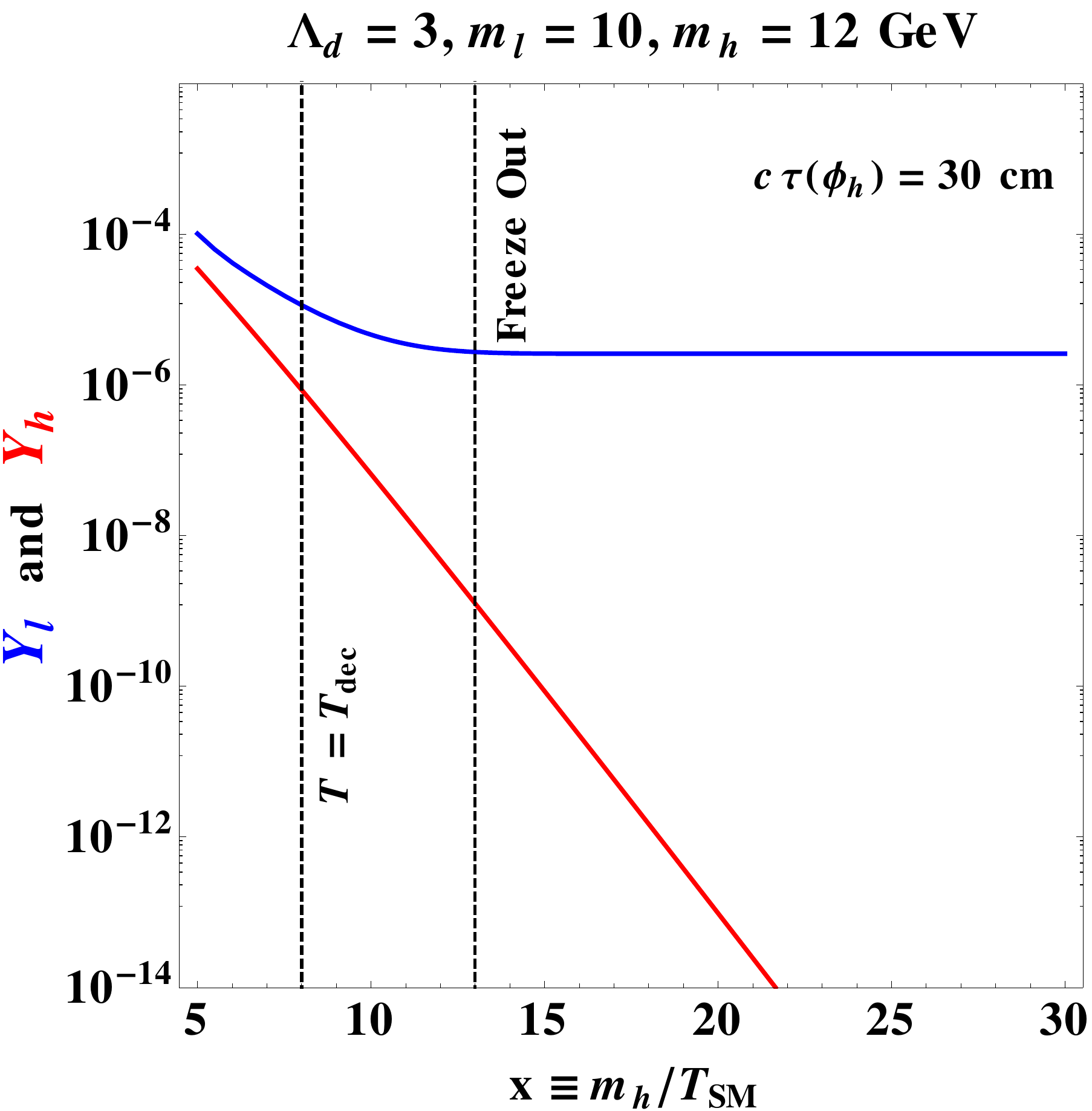}\qquad
\includegraphics[height=6cm]{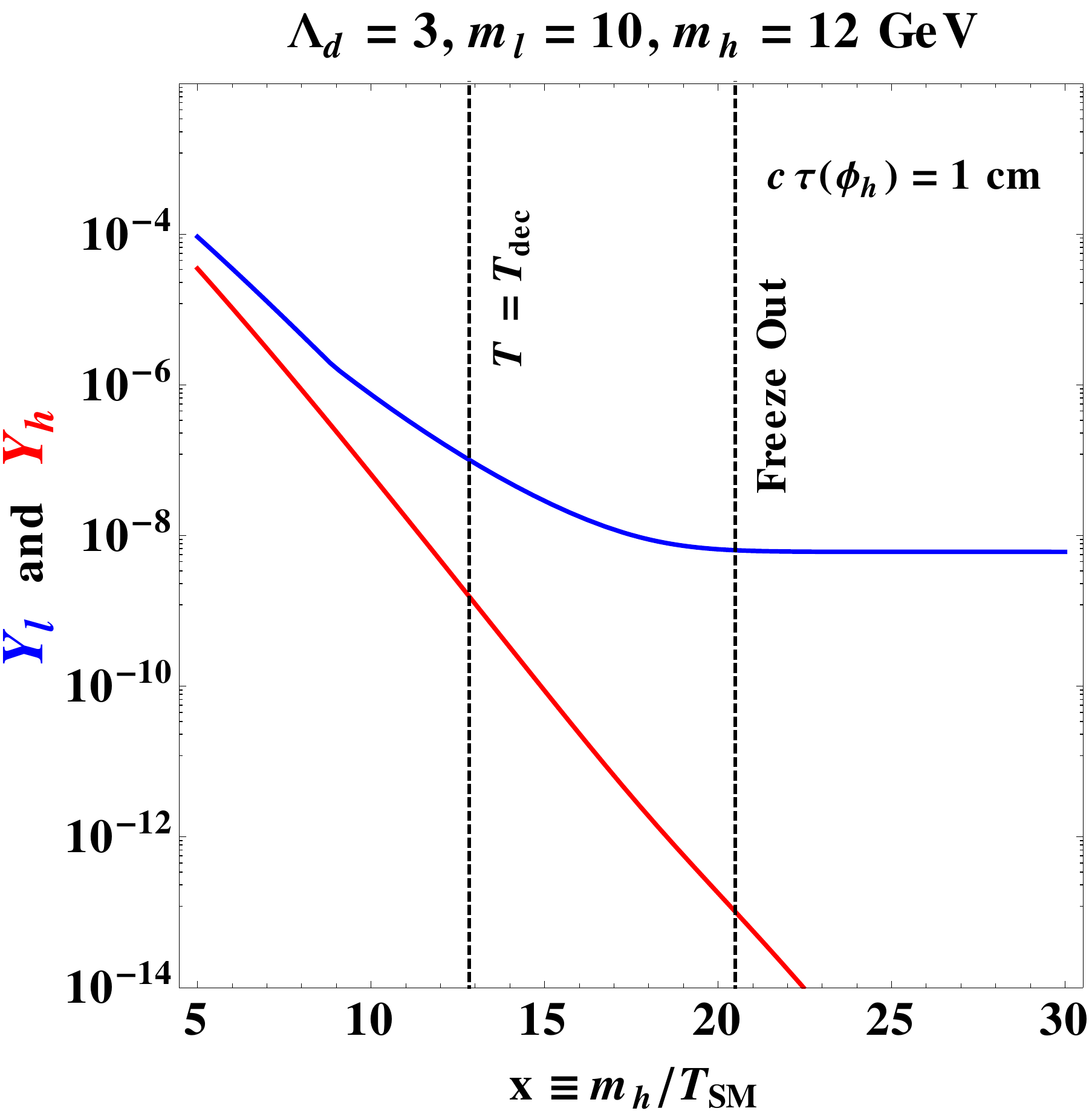}

\caption{
Evolution of $Y_{l,h}$ in $x \equiv m_h/T_{\rm SM}$ for different dark hadron masses and $\phi_h$ lifetimes. The blue and red curves stand for $Y_l$ and $Y_h$ separately. The vertical lines at smaller $x$ values show the temperature of thermal decoupling between the hidden and SM sectors. The vertical lines at larger $x$ values show the temperature of $\eta_d$ freeze out. We assume $Y_h=Y_l$ for the initial condition.
}
\label{fig:Second_Results}
\end{figure}

The energy transfer rate between the SM and dark sector can come from the decay-inverse decay process $\phi_{h}\leftrightarrow$ SM, and the rate is written as
\begin{align}
\frac{dE}{dt}\Big|_{\ell(k_1)\ell(k_2)\to \phi_h(p)} &= \int  d \Omega_{k1} d \Omega_{k2} d \Omega_{p} f_{k_1} f_{k_2} \frac{|\mathcal{M}|^2}{4 E_{k_1}E_{k_2}} (E_{k_1}+E_{k_2})(2\pi)^4 \delta^4(k_1+k_2+p),\nonumber
\\
&\simeq(2\pi)^{-2}\iint \frac{d^3 k}{2 E_k} f_k^2 2E_k (16 \pi \,m_h \Gamma_{\phi_h}) \delta (E_k-m_h/2),\nonumber\\
&\simeq \frac{4}{\pi} \Gamma_{\phi_h} m_h e^{-m_h/T}.\label{eqn:energy_exchange_1}
\end{align}
In scenarios with a Higgs portal coupling, dark hadron and SM fermions can scatter elastically via the Higgs exchange. This introduces additional energy transfer between the two sectors~\cite{Craig:2016lyx}
\begin{align}
\label{eqn:energy_exchange_2}
\frac{dE}{dt}_{q_d(p_1)f(k_1)\to q_d(p_2)f(k_2)}\simeq  y_{q_d}\,y_b \frac{m_l T^5 }{8 \pi^3 m_h^4} e^{-(m_b+m_l)/T}.
\end{align}
Depending on the hadron mass and dark yukawa coupling, this energy transfer rate can be comparable to Eq.~(\ref{eqn:energy_exchange_1}). However, since Eq.~(\ref{eqn:energy_exchange_2}) depends on the assumptions of dark yukawa coupling, and the same Boltzmann suppression makes the $T_{dec}$ from Eq.~(\ref{eqn:energy_exchange_2}) to be similar to Eq.~(\ref{eqn:energy_exchange_1}), we will only include Eq.~(\ref{eqn:energy_exchange_1}) from the $\phi_h$ decay-inverse decay when solving the temperature evolution in Eq.~(\ref{eq:Tdec2}).

Before solving Eqs.~(\ref{eqn:boltz_eta_1},~\ref{eqn:boltz_eta_2}) numerically, let us first seek some analytical understanding of the freeze-out temperature $T_{FO}$ and the relic abundance $Y_l$ from the conversion/decay process. Since the freeze out process relies on the $\phi_h$ decay, we focus on the temperature scale when $\Gamma_h \gg H$. In this case, the decay terms in Eq.~(\ref{eqn:boltz_eta_1}) keeps the comoving number density $Y_h$ to thermal distribution\footnote{Even though the energy transfer from Eq.~(\ref{eqn:energy_exchange_1}) is too small to keep the whole $\phi_{h,l}$ sector in kinetic equilibrium with the SM, the exponentially suppressed inverse decay can still keep the exponentially suppressed $Y_h$ number before $T_{FO}$.}
\begin{equation}\label{eq:Yhnum}
Y_h \approx Y_h^{\text{eq}}(T)\approx\exp\left(-\frac{m_h}{T}\right).
\end{equation}
Near the freeze out time, Eq.~(\ref{eqn:boltz_eta_2}) can be approximated as
\begin{equation}\label{eq:approxYl}
\left(\frac{Y_l}{x}\right)^{-1}\frac{dY_l}{dx}\sim \frac{T^3\langle\sigma_{-h}v\rangle}{H(x)}e^{-\frac{m_h}{T}}\left[1-e^{-(\frac{\Delta m}{\hat{T}}-\frac{m_h}{T})}Y_l\right].
\end{equation}
Here we neglect the sub-leading contribution from the $\langle\sigma_{\pm 2h}v\rangle$ terms that carry either an additional $e^{-\frac{m_h}{T}}$ or $e^{-\frac{\Delta m}{\hat{T}}}$ suppression.

$Y_l$ freezes out when its temperature evolution is close to the thermal distribution
\begin{equation}\label{eq:Ylnum}
Y_{l,\,FO}\approx \exp\left(\frac{\Delta m}{\hat{T}}-\frac{m_{h}}{T}\right)\approx\exp\left(\frac{\Delta m\,T_{dec}}{T^2}-\frac{m_{h}}{T}\right), 
\end{equation}
where $T_{dec}$ can be estimated by solving Eq.~(\ref{eq:Tdec}). As $Y_l$ drops when the temperature is getting close to $T_{FO}$, its value is asymptotically close to Eq.~(\ref{eq:Ylnum}). Since $Y_l$ freezes out when having vanishing derivative in $x=m_h/T$, we can estimate $x_{FO}$ by solving
\begin{equation}\label{eq:FOestimate}
\frac{d \log Y_l}{d x} \approx \frac{d \log Y_{l,FO}}{d x} \approx \frac{2\,\Delta m\,T_{\text{dec}}}{m_h^2}\,x_{FO} -1= 0.
\end{equation}
This gives $x_{FO}\approx m_h^2/(2\Delta m\,T_{dec})$. However, since the pre-factor in front of the square bracket in Eq.~(\ref{eq:approxYl}) can be smaller than $1$ depending on the parameter choice, freeze-out can happen at a slightly smaller $x_{FO}$ than this rough estimate. For the dark hadron parameters studied in this work, numerical solution gives $x_{FO}= (1-c)\,m_h^2/(2\Delta m\,T_{dec})$ with $0\leq c\lsim 0.7$. The relic abundance of $Y_l$ before $\phi_l$ decays can be estimated using Eq.~(\ref{eq:Ylnum})
\begin{equation}\label{eq:Ylestimate}
Y_{l,FO} = \exp\bigg[-\left(\frac{1-c^2}{4}\right)\frac{m_h^2}{\Delta m\,T_{dec}}\bigg].
\end{equation}
We should emphasize again the expression is derived by assuming $m_{h,l}\gg T_{dec}$ and $\Gamma_h\gg H$. The result shows that $\phi_l$ gets a larger relic density either if the fractional mass gap $(\Delta m/m_h)$ is larger, or the decoupling temperature $T_{dec}$ is higher. The lower $T_{dec}$ corresponds to a larger $\Gamma_h$ in the energy transfer rate Eq.~(\ref{eqn:energy_exchange_1}), and the upper bound on $Y_{l,FO}$ from cosmological constraints sets a lower bound on $\Gamma_h$.

In Fig.~\ref{fig:Second_Results}, we show four examples of the $Y_{l,h}$ evolution under two assumptions of dark hadron mass and $c\tau_{\phi_h}$. From the numerical results, we check that before $\phi_l$ freezes out, the analytical estimates in Eqs.~(\ref{eq:Yhnum}, \ref{eq:Ylnum}) match very well with the numerical results. As expected, a longer $c\tau(h)$ introduces a higher $T_{\text{dec}}$. 
 A larger mass gap ratio $(\Delta m/m_{h})$ 
makes the freeze out happen at a smaller $x$ and leads to a larger $Y_{l,FO}$. 

\begin{figure}
\centering
\includegraphics[scale=0.32]{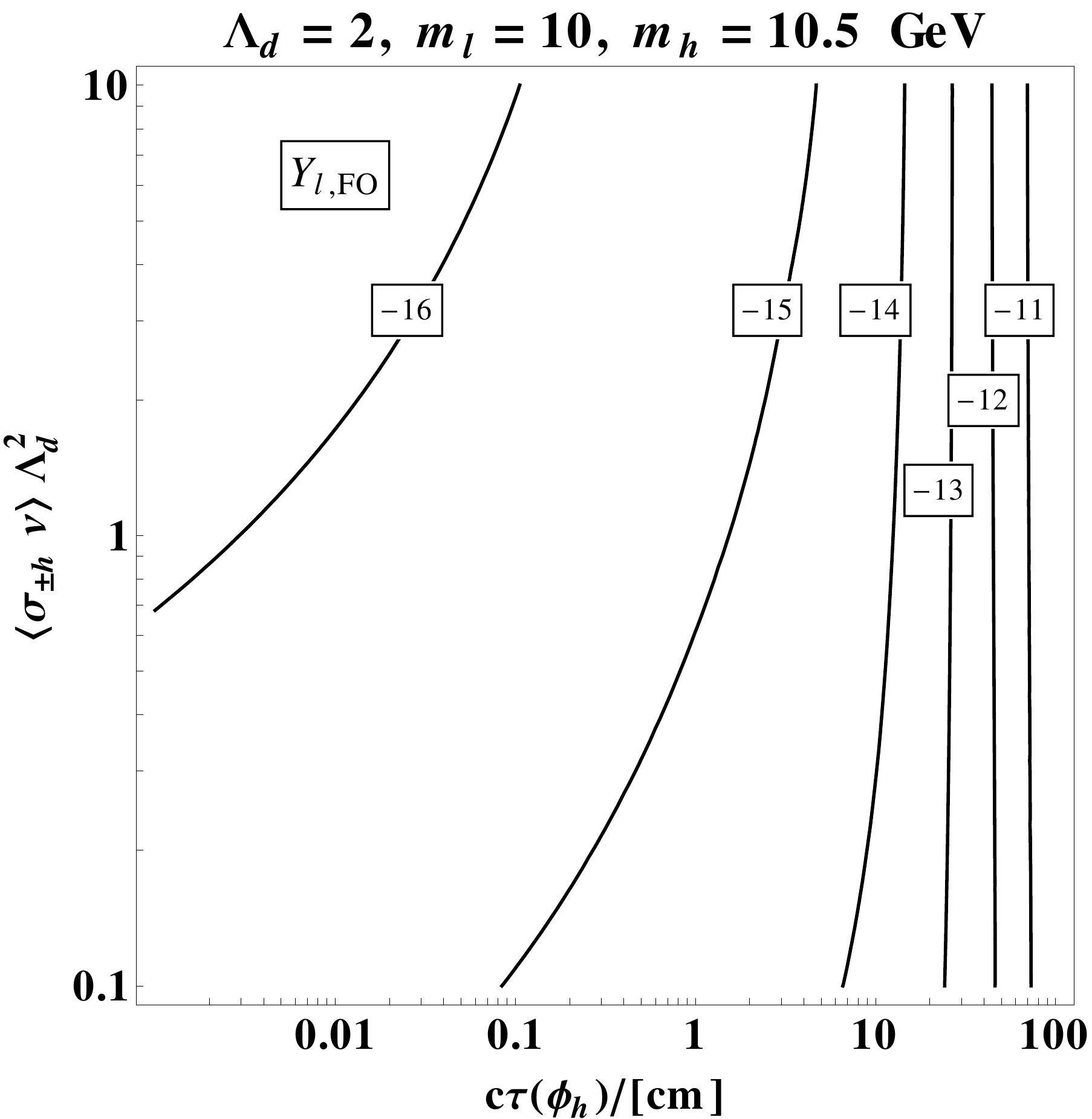}\qquad
\includegraphics[scale=0.32]{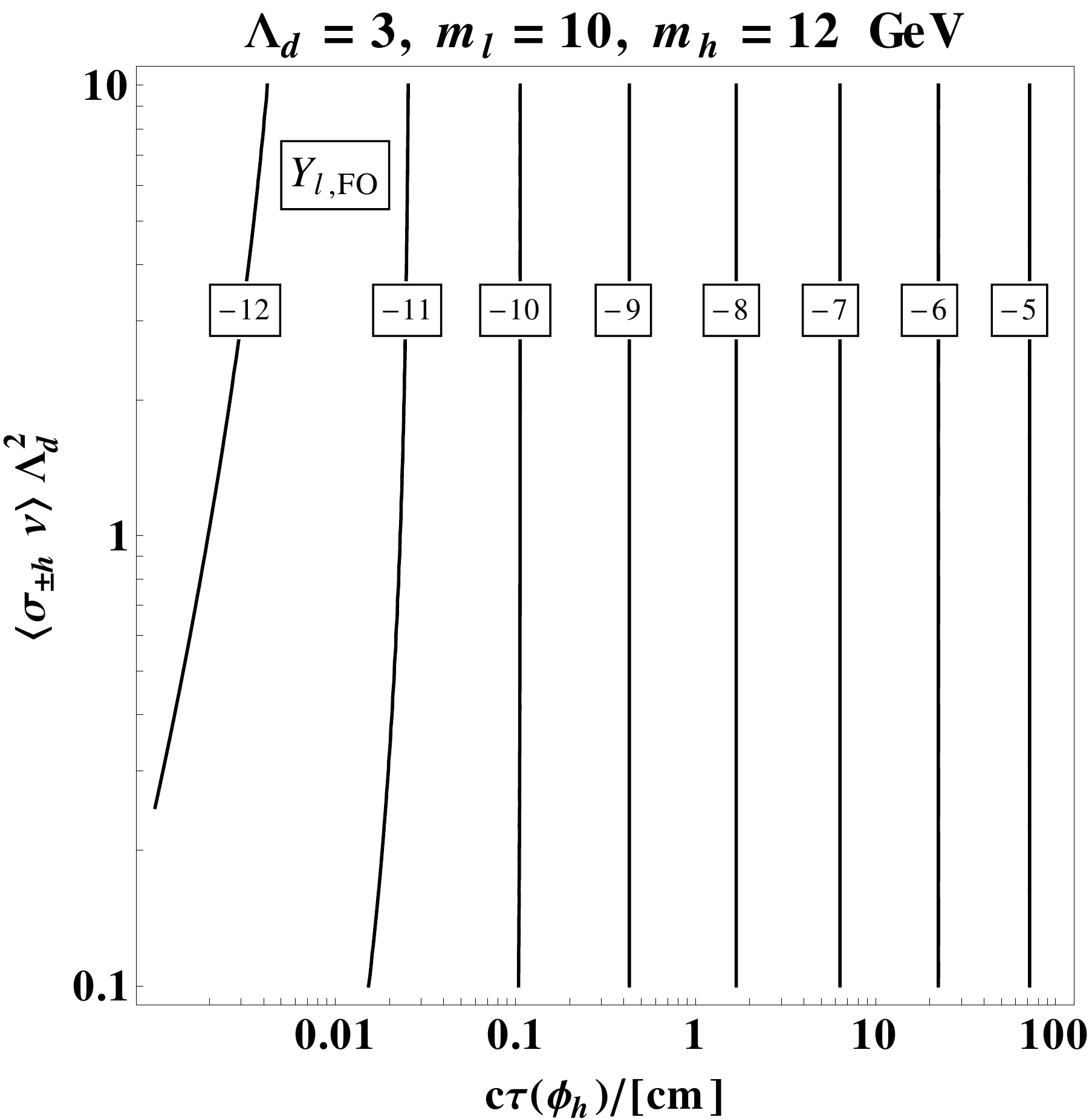}
\caption{The result of $Y_l$ for different $\langle \sigma_{\pm h}v\rangle\Lambda_d^2$ assumptions. For the lifetime bound we are interested in, $c\tau_{\phi_h}\gsim1$ cm, and the result is insensitive to $\langle \sigma_{\pm h}v\rangle\Lambda_d^2$. }
\label{fig:Scanning}
\end{figure}

When solving the Boltzmann Eqs.~(\ref{eqn:boltz_eta_1}, \ref{eqn:boltz_eta_2}), the size of dark hadron conversion depends on details of the dark QCD coupling. However, since the relic $\phi_l$ abundance is mainly determined by $(\Delta m/m_h)$ and $T_{dec}$, and $T_{dec}$ is insensitive to the hadronic cross sections, the final $Y_l$ is quite insensitive to the $\langle\sigma_{\pm h}v\rangle$ value. This is not true, however, if the inverse decay of $\phi_h$ is highly efficient to keep the SM-dark sector thermal equilibrium down to a very low temperature. The thermal equilibrium makes dark sector warmer and allows $\phi_l$ to keep converting into $\phi_h$ until its number density times $\langle\sigma_{\pm (2)h}v\rangle$ is smaller than Hubble. In this case $Y_l$ is more sensitive to $\langle\sigma_{\pm h}v\rangle$. In Fig.~\ref{fig:Scanning}, we show two examples of the $Y_l$ contours with different dark hadron masses. When $c\tau_{\phi_h}\ll1$ cm, the relic $Y_l$ becomes sensitive to the ratio between $\langle\sigma_{\pm h}v\rangle$ and $\Lambda_d^{-2}$. Since we are interested in cosmological bounds relating to $c\tau_{\phi_h}\gsim1$ cm, the result is less sensitive to the hadronic cross sections. We therefore choose $\langle\sigma_{\pm (2)h}v\rangle=\Lambda_d^{-2}$ to reduce the number of parameters in the rest of the analysis.

\section{Upper bounds on the dark hadron lifetime}\label{sec:3}
If $\phi_l$ later decays into SM hadrons or charged leptons, the BBN and CMB constraints set an upper bounds on $Y_{l,FO}$ right before the decay. In this work, we adapt the BBN bound in Fig.~11 of Ref.~\cite{Kawasaki:2017bqm} when setting the $Y_{l,FO}$ constraint. The BBN bound requires $m_l\,Y_{l,FO}\lsim 10^{-9}$ GeV for $\tau_{\phi_l}> 1\,(100)$ sec if $\phi_l$ decays into SM hadrons (charged leptons). As discussed below Eq.~(\ref{eq:Ylestimate}), these upper bounds on $Y_{l,FO}$ set upper bounds on $T_{dec}$, which correspond to minimum decay rates of $\phi_h$. When calculating the $c\tau_{\phi_h}$ bounds, however, we solve the Boltzmann Eqs.~(\ref{eqn:boltz_eta_1},\,\ref{eqn:boltz_eta_2}) numerically instead of using the approximate form in Eq.~(\ref{eq:Ylestimate}). 

To illustrate the idea, we discuss two benchmark scenarios of the confining hidden valley in the following sub-sections. We first consider a dark sector that couples to the SM sector through a kinetic mixing between the dark photon and the SM hypercharge gauge field. In this scenario, a pseudo-scalar meson ($\eta_d$) is the lightest dark particle, and the BBN constraint sets an upper bound on the lifetime of the lightest vector meson ($\omega_d$). For the second scenario, we consider the two sectors couple with each other through a Higgs portal coupling, and the CP-symmetry is unbroken in the hidden sector. The pseudo-scalar meson ($\eta_d$) is still the lightest dark particle, and the constraint on DM density sets a lifetime bound on the lightest scalar meson ($\chi_d$) or the lightest scalar glueball ($\widetilde{G}_{0^{++}}$).

\subsection{Photon portal scenario}

We consider the lightest dark hadron to be a pseudo-scalar meson ($\eta_d=\phi_l$), and the heavier hadron to be a vector meson ($\omega_d=\phi_h$). To simplify the assumption of dark hadron spectrum, we consider heavy quark scenarios where dark quarks are heavier than the confinement scale, $m_{q_d}>\Lambda_d$. In this case, $\eta_d$ is the lightest meson, and $\omega_d$ is the heavier meson state in the triplet hyperfine state with a mass splitting to be some order one fraction of $\Lambda_d$~\cite{Nussinov:1999sx}. We only consider scenarios with $\Lambda_d>\frac{1}{7}m_{\omega_d}$, so $\omega_d$'s cannot annihilate into scalar glueballs~\cite{Chen:2005mg}. When having different assumptions of hadron masses, one can follow the same analysis in this work to obtain different lifetime constraints. 

We assume the only coupling between the SM and dark particles comes from a kinetic mixing between a dark photon $Z_d$ and the SM U$(1)_Y$ gauge field
\begin{equation}\label{eq:kinmix}
\mathcal{L}_{\rm Dark}\supset \bar{q}_d(i\slashed\partial +g_{d}\slashed Z_d)q_d+\frac{m_{Z_d}^2}{2} Z_{d,\mu}Z^{\mu}_d+\frac{\epsilon}{2}B_{\mu\nu}F^{\mu\nu}_d.
\end{equation}
In this work, we consider scenarios that have above above GeV-scale dark hadron mass and dark photon mass $m_{Z_d}>2m_{\eta_d}$. In this case, the pseudo-scalar cannot simply decay into a pair of dark photons before the BBN, and the $\eta_d$ abundance is set both by the $\eta_d$-$\omega_d$ conversion and the $\omega_d$ decay. The heavier dark photon assumption is also motivated by the stronger collider and astrophysical constraints on dark photons below GeV-scale. 

Since the pseudo-scalar $\eta_d$ cannot decay into SM particles through a single gauge boson\footnote{The inner product between the derivative coupling of the pseudo-scalar and the kinetic mixing operator $\sim (p^2g^{\mu\nu}-p^{\mu}p^{\nu})$ vanishes~\cite{Essig:2009nc}.}, $\eta_d$ only decays either into four SM fermions via two off-shell $Z_d$'s, or into two SM fermions through a loop level process. The corresponding decay rates can be estimated as~\cite{Hochberg:2018vdo}
\begin{equation}
\Gamma_{\eta_d \to 4f} \approx \frac{9 \alpha_d^2m_{\eta_d}^3}{8\pi^3 f_{\eta_d}^2} \bigg[ \frac{\alpha}{2\pi} \epsilon^2  \bigg(\frac{m_{\eta_d}}{2 m_{Z_d}}\bigg)^4 \bigg]^2,\quad\Gamma_{\eta_d \to f\bar{f}} \approx \frac{9 \alpha_d^2m_{\eta_d}^3}{8\pi^3 f_{\eta_d}^2} \bigg[ \frac{\alpha}{2\pi} \epsilon^2 \bigg(\frac{m_f}{m_{\eta_d}}\bigg)^2 \bigg(\frac{m_{\eta_d}}{m_{Z_d}}\bigg)^2 \bigg]^2, 
\label{eqn:etalifetime}
\end{equation}
and we take the decay constant $f_{\eta_d}=\Lambda_d$ and dark photon coupling $\alpha_d=\alpha$ in the estimate. For GeV-scale $(m_{\eta_d},\,\Lambda_d)$ and $\epsilon\sim 10^{-2}$, the size of $\tau_{\eta_d}$ can be easily above $100$ sec if $m_{Z_d}$ is few times heavier than $m_{\eta_d}$. Such a decay is tightly constrained by the BBN bounds.  Depending on the ultraviolet completion model, it is also possible to make the pseudo-scalar a DM particle~\cite{Berlin:2018tvf}, and the lifetime bound in this case comes from requiring $\Omega_{\eta_d}<\Omega_{\rm DM}$. 

Different from the pseudo-scalar meson, the vector bound state $\omega_d$ has a much faster decay into SM fermions via a single off-shell $Z_d$, which is the dominant process to release the dark hadron energy into SM. We take the following lifetime expression for $\omega_d$ decaying into SM fermions based on the estimation in Ref.~\cite{Cheng:2015buv} assuming $m_{q_d}\gg\Lambda_d$
\begin{equation}\label{eq:ctauphoton}
c\tau_{\omega_d}\approx 30\,{\rm cm}\left(\frac{10\,{\rm GeV}}{m_{\eta_d}}\right)^3\left(\frac{m_{Z_d}}{100\,{\rm GeV}}\right)^4\left(\frac{10^{-3}}{\epsilon}\right)^2\left(\frac{1\,{\rm GeV}}{\Lambda_d}\right)^2.
\end{equation}
The expression is derived using the quirky bound state physics~\cite{Kang:2008ea}. 

\label{sec:photonportal}

\begin{figure}
\centering
\includegraphics[scale=0.36]{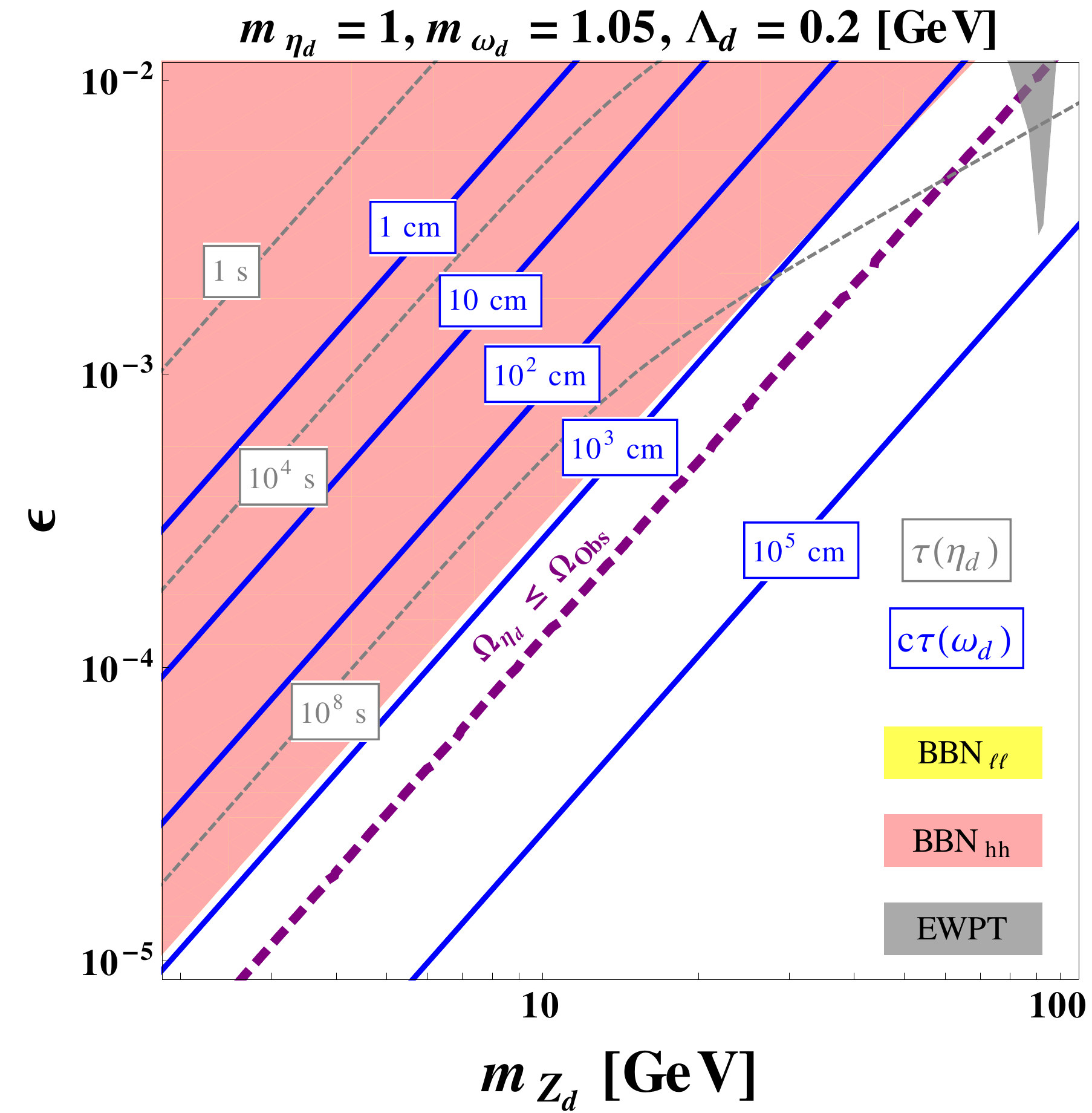}\qquad
\includegraphics[scale=0.36]{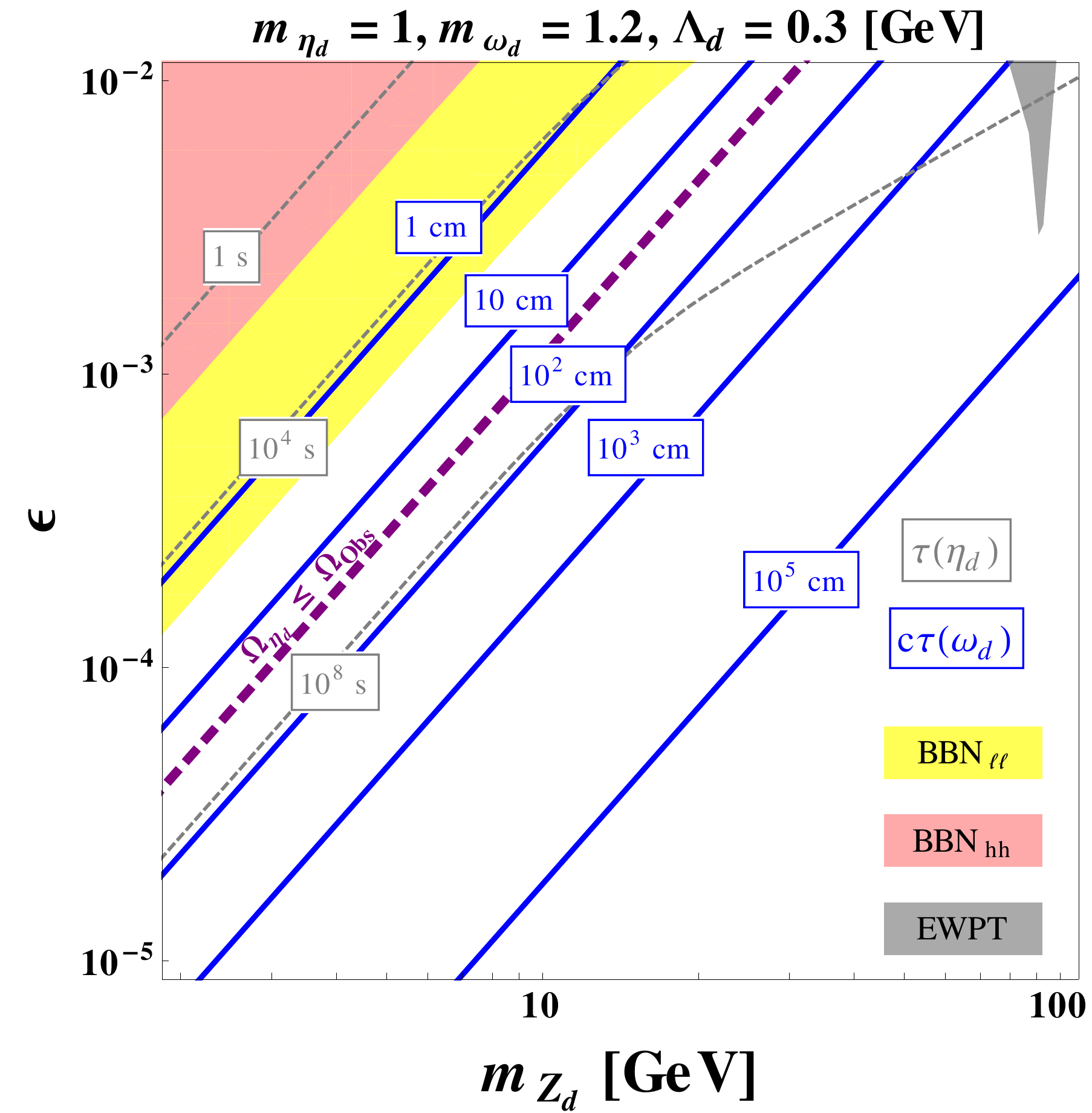}
\\\vspace{1em}
\includegraphics[scale=0.36]{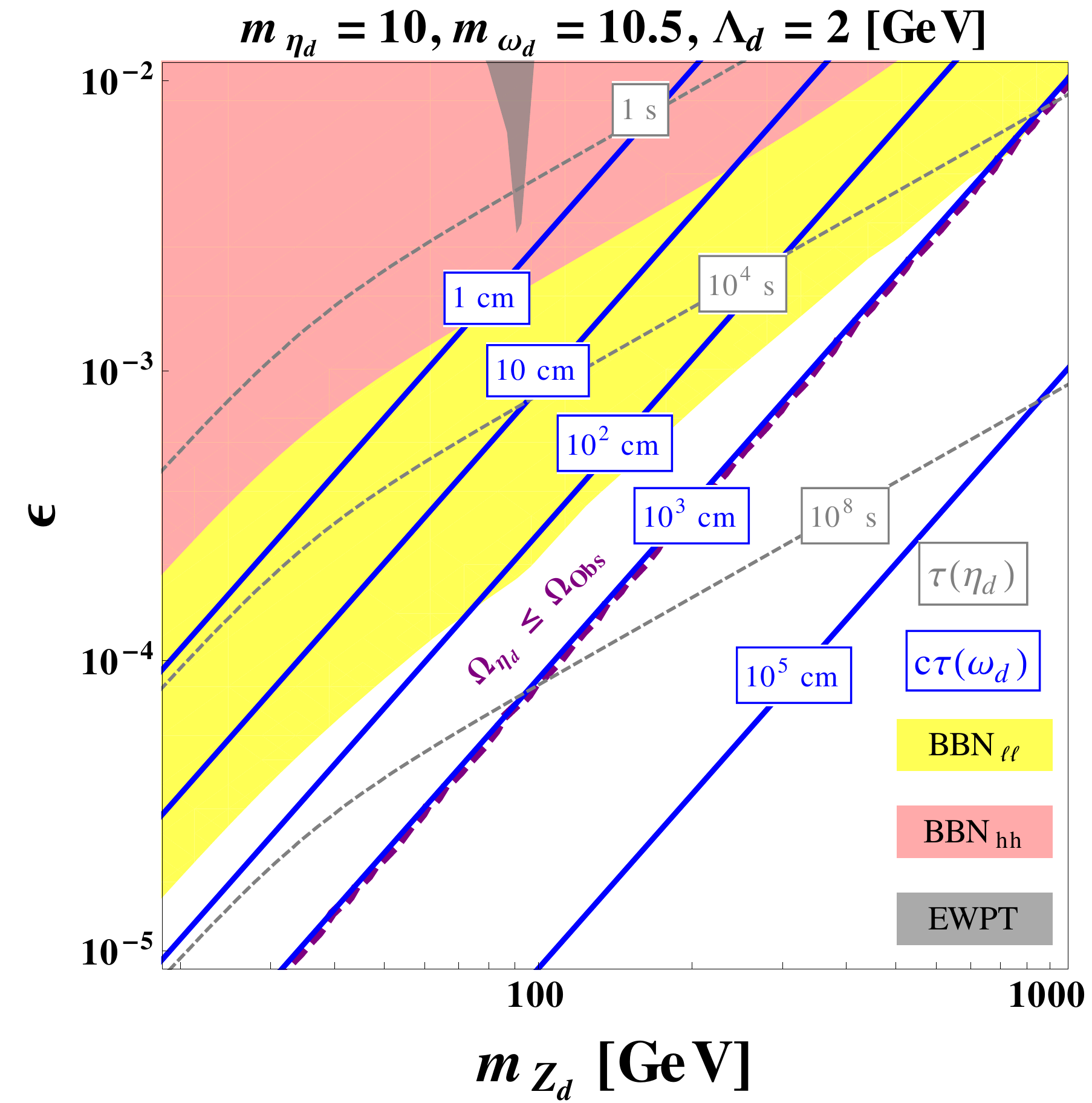}\qquad
\includegraphics[scale=0.36]{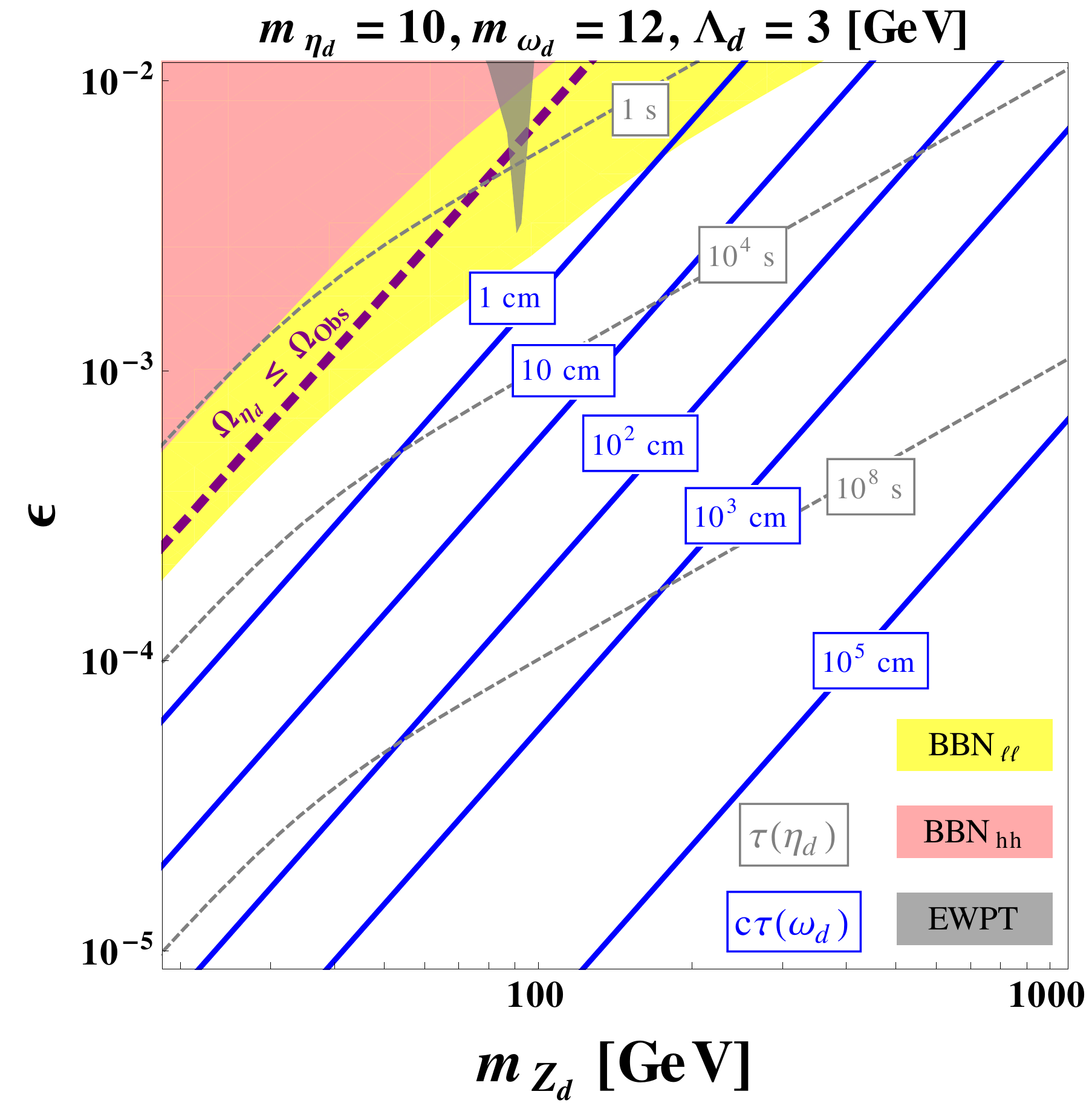}
\caption{BBN bounds on the photon portal scenario with two choices of $m_{\eta_d}$ and $(m_{\omega_d}/m_{\eta_d})$ ratio. We consider scenarios with $m_{q_d}>\Lambda_d$. The pink (yellow) regions give the allowed dark photon mass and mixing from the BBN constraints if $Z_d$ decays dominantly into SM hadrons (leptons). As a reference, the dashed purple line corresponds to the parameters when the $\eta_d$ abundance equals the observed DM abundance before it decays. The gray shaded region is the parameter space excluded by the current electroweak precision test data. Blue (gray) curves give lifetimes of $\omega_d$ ($\eta_d$) from the estimates in Eq.~(\ref{eq:ctauphoton}, \ref{eqn:etalifetime}). Notice that the range of $m_{Z_d}$ values are different between the upper and lower plots.}
\label{fig:DP1}
\end{figure}

We consider the following conversions and decay
\begin{equation}
\pm2h:\,\,\eta_d\,\eta_d\leftrightarrow\omega_d\,\omega_d,\qquad\pm h:\,\,\eta_d\,\eta_d\leftrightarrow\eta_d\,\omega_d,\qquad\omega_d\to f\bar{f}\,\,\,{\rm in\,\,\,SM}.
\end{equation}
The coupling between dark and SM sectors comes from the decay-inverse decay of $\omega_d$. The $\pm h$ process requires the presence of more than one dark quark flavor, so the three $\eta_d$'s in the conversion need to carry different dark charges. When solving the relic abundance of pseudo-scalar mesons, since the different $\eta_d$'s are in chemical equilibrium, we simply consider the $\phi_l$ density in the Boltzmann equations to be the sum of different $\eta_d$ densities. If the neutral $\eta_d$ later decays into SM particles, the number of charged $\eta_d$'s also drops from annihilating into the neutral state. We did a numerical study by only keeping the $\pm 2h$ process in the equations for the single $q_d$-flavor scenario, and the resulting lifetime constraint changes mildly. Since $\omega_d$ has three spin degrees of freedom, we take $Y_{\omega_d}=3Y_{\eta_d}$ as the initial condition at high temperature.

Once obtaining the relic abundance of $\eta_d$, we use it to set upper bounds on $c\tau_{\omega_d}$ from the BBN constraint. In Fig.~\ref{fig:DP1}, we show the lifetime constraints in two $m_{\eta_d}$ values (upper and lower plots) and two $m_{\omega_d}$ over $m_{\eta_d}$ ratios (left and right plots).
The $\omega_d$ and $\eta_d$ lifetimes are shown as the blue and dashed-gray curves. The BBN constraints on $m_{\eta_d}Y_{\eta_d}$ depends on $\eta_d$ lifetimes, and the size of $Y_{\eta_d}$ before the decay is determined by $c\tau_{\omega_d}$ (blue lines). The $\Omega_{\eta_d}=\Omega_{\rm DM}$ line (dashed-purple) is parallel to the blue lines and corresponds to a smaller $c\tau_{\omega_d}$ in the right plots comparing to the left plots. This shows how the earlier termination of the $+(2)h$ process due to a larger $(m_{\omega_d}/m_{\eta_d})$ ratio requires a faster $\omega_d$ decay to obtain the same $\eta_d$ abundance. The same behavior happens for the BBN bounds, where the pink and yellow shaded region corresponds to smaller $c\tau_{\omega_d}$ in the right plots. 

The BBN constraint requires $\Omega_{\eta_d}/\Omega_{\rm DM}\lsim 10^{-4}$ if $\eta_d$ decays in the early universe between $10^6-10^{23}$ sec, and $\Omega_{\eta_d}/\Omega_{\rm DM}\lsim 1$ if $\eta_d$ decays between $1-100$ sec. This is why all the BBN covered regions are above the $\Omega_{\eta_d}=\Omega_{\rm DM}$ lines (dashed-purple) unless $\eta_d$ decays within $100$ sec as in the lower right plot. Since the BBN bounds constrain the energy density of $\eta_d$, the heavier $m_{\eta_d}$ in the lower plots get tighter constraints comparing to the $c\tau_{\omega_d}$ values. Most of the allowed parameter space requires $c\tau_{\omega_d}<1$ m, and having a larger $(m_{\omega_d}/m_{\eta_d})$ ratio gets even stronger bounds. Once being produced at colliders, the long-lived $\omega_d$ needs to have a fast enough decay inside particle detectors.

The $Z_d$ we consider decays dominantly into dark hadrons, and dark photon constraints that require visible $Z_d$ decays do not apply. However, once we produce $Z_d$ or dark quarks in a collider experiment, $\omega_d$ can be formed and generate long-lived particle signatures. In Sec.~\ref{sec:search}, we discuss an example of the $\omega_d$ signature from the exotic $Z$ boson decay and show how does the BBN constraint narrow down the parameter space for collider searches.

\subsection{Higgs portal scenario}
\label{sec:Higgsportal}

We consider scenarios where dark quarks only couple to SM particles through a Higgs portal coupling
\begin{equation}\label{eq:higgsportal}
\mathcal{L}_{\rm Dark}\supset -\frac{y_d}{\sqrt{2}} \frac{v}{f}\,h\,\bar{q}_dq_d\,,
\end{equation}
where $h$ and $v$ are the SM Higgs boson and its vacuum expectation value (VEV). $f$ can be considered as a dark Higgs VEV, and the mixing coupling exists, e.g., in the Twin Higgs model~\cite{Chacko:2005pe,Craig:2015pha} that solves the little hierarchy problem. We assume the CP-symmetry is unbroken in the hidden sector. Since the Higgs portal coupling connects CP even states, the pseudo-scalar meson $\eta_d$ does not decay into SM particles through this coupling. Being the lightest state of dark hadrons, $\eta_d$ becomes a DM component and has its relic abundance constrained by the observed $\Omega_{\rm DM}h^2$. 

For the mass and coupling we consider, the annihilation process $\eta_d\eta_d\to$ SM particles through the Higgs portal coupling is inefficient to obtain $\Omega_{\eta_d}<\Omega_{\rm DM}$. When focusing on GeV-scale mesons, the $3\to 2$ annihilation of $\eta_d$ is also too slow to lower the $\eta_d$ abundance~\cite{Hochberg:2014dra}. We therefore have to rely on the $\eta_d$ conversion into heavier hadrons to suppress the relic abundance. Depending on the mass splitting to $\eta_d$, the heavier hadron $\phi_h$ can be either be the scalar meson $\chi_d$ or the scalar glueball $\widetilde{G}_{0^{++}}$. The conversion and the decay we consider are
\begin{equation}\label{eq:Higgsprocess}
\pm2h:\,\,\eta_d\eta_d\leftrightarrow\chi_d\chi_d\,\,{\rm or}\,\,\widetilde{G}_{0^{++}}\widetilde{G}_{0^{++}},\,\qquad \chi_d\,\,{\rm or}\,\,\widetilde{G}_{0^{++}}\to f\bar{f}\,\,\,{\rm in\,\,\,SM}.
\end{equation}
Due to the parity conservation, the $\pm h$ process discussed in Sec.~\ref{sec:2} requires the participation of other mesons such as $\eta_d\eta_d\leftrightarrow\chi_d\omega_d$. Unless $\omega_d$ is much lighter than $\chi_d$, this additional process is not more efficient than the $\pm 2h$ process we consider. We therefore only include the $\pm2h$ process and estimate the lifetime bounds on $\chi_d$ or $\widetilde{G}_{0^{++}}$. When solving Boltzmann equations, we take $Y_{\eta_d}=Y_{\chi_d}$ or $Y_{\widetilde{G}}$ as the initial condition at high temperature. To simplify the discussion, we assume following relations between $(\Lambda_d,f)$ and dark particle masses
\begin{equation}\label{eq:mesonm}
m_{\chi_d}\equiv 2(m_{q_d}+\Lambda_d),\quad m_{q_d}=\frac{y_d}{\sqrt{2}}\,f,\quad m_{\widetilde{G}_{0^{++}}}=6.9\,\Lambda_d\,.
\end{equation}
The choice of $m_{\chi_d}$ is motivated by the SM $\chi_c$ ($c\bar{c}$) mass, and the glueball mass comes from the lattice calculation in a pure Yang-Mills theory~\cite{Chen:2005mg}. 

\begin{figure}
\centering
\includegraphics[scale=0.36]{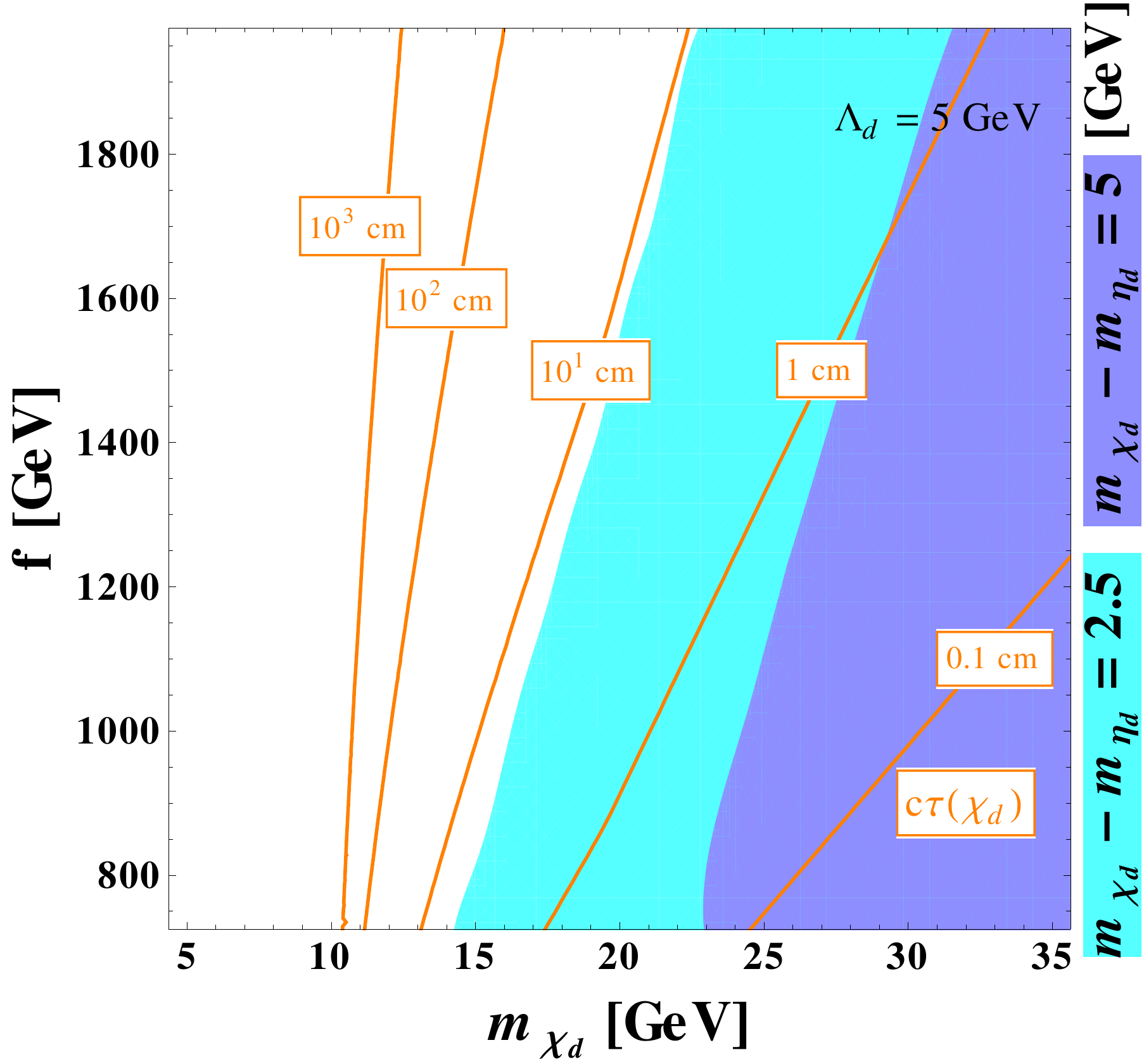}\qquad\includegraphics[scale=0.36]{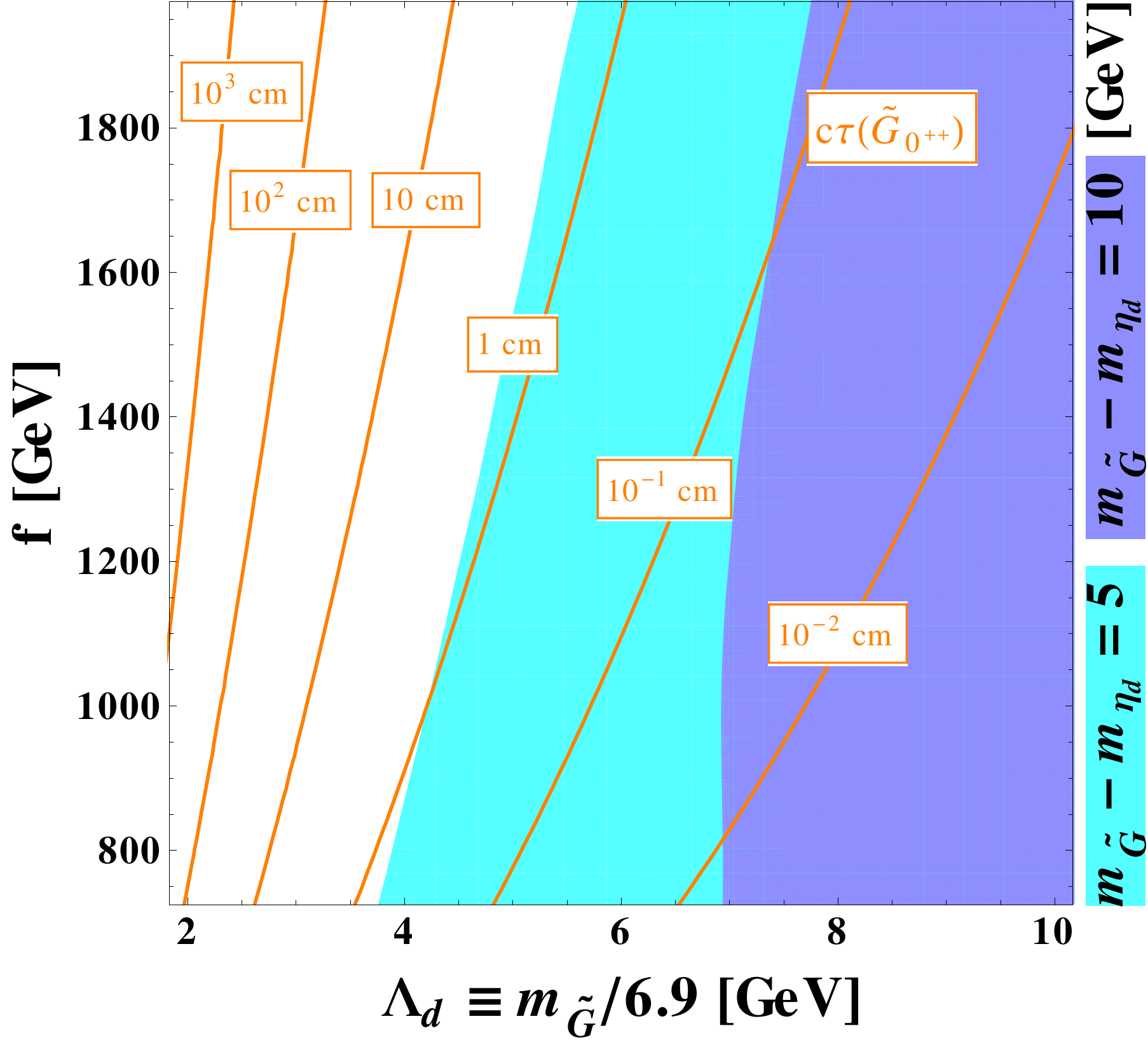}
\caption{\emph{Left}: Constraints on the scalar meson lifetime in the Higgs portal scenario derived by requiring $\Omega_{\eta_d}<\Omega_{\rm DM}$. The light and dark blue regions both extend to $m_{\chi_d}=35$ GeV before $\chi_d$ becomes heavier than the scalar glueball. \emph{Right}: Constraints on the glueball lifetime assuming $m_{\widetilde{G}_{0^{++}}}<m_{\chi_d}$. The dark blue region is laid on top of the light blue region. The mass relation Eq.~(\ref{eq:mesonm}) is assumed in these plots.}
\label{fig:chi}
\end{figure}

If $m_{\chi_d}\ll m_{\widetilde{G}_{0^{++}}}$, the heavier meson to consider in Eq.~(\ref{eq:Higgsprocess}) is $\chi_d$. We use the lifetime estimated in Ref.~\cite{Cheng:2015buv} 
\begin{eqnarray}
c\tau_{\chi_d}&\approx& 4.8 \bigg(\frac{m_b}{m_{q_d}} \bigg)^4 \bigg( \frac{f}{1~\text{TeV}}\bigg)^4 \bigg(\frac{5~\text{GeV}}{\Lambda_d}\bigg)^3~\text{cm},~~(m_{q_d} \geq \Lambda_d)
\\
c\tau_{\chi_d}&\approx& 9.0 \bigg(\frac{m_b}{m_{q_d}} \bigg)^2 \bigg( \frac{f}{1~\text{TeV}}\bigg)^4 \bigg(\frac{5~\text{GeV}}{\Lambda_d}\bigg)^5  \bigg(\frac{m_{\chi_d}}{2\Lambda_d} \bigg)^{-2}~\text{cm},~~(m_{q_d} < \Lambda_d)
\end{eqnarray} 
and the hadronic conversion cross section $\langle\sigma_{\pm 2h} v\rangle=\Lambda_d^{-2}$ in the Boltzmann equations. As discussed in the end of Sec.~\ref{sec:2}, the lifetime constraint is quite insensitive to $\langle\sigma_{\pm 2h} v\rangle$. We show an example of the $\chi_d$ lifetime bound in Fig.~\ref{fig:chi} (left) by assuming $\Lambda_d=5$ GeV. When the mass difference between $\eta_d$ and $\chi_d$ is larger than $0.5\Lambda_d$, the observed $\Omega_{\rm DM}h^2$ value requires $c\tau_{\chi_d}\leq 10$ cm. $\eta_d$ can also be the dominant DM component when $c\tau_{\chi_d}$ is close to this lifetime bound. 

If $m_{\widetilde{G}_{0^{++}}}<m_{\chi_d}$, the heavier meson to consider in Eq.~(\ref{eq:Higgsprocess}) is the scalar glueball. We use the glueball lifetime derived in Ref.~\cite{Craig:2015pha} for the Fraternal Twin Higgs model to solve the $\eta_d$ abundance

\begin{equation}
\Gamma_{\widetilde{G}_{0^{++}}\to {\rm SM}}=\bigg[ \frac{3.06\,\hat{\alpha}_s m_{\tilde{G}_{0^{++}}}^3 v}{24\pi^2 f^2 (m_h^2-m_{\tilde{G}_{0^{++}}}^2)} \bigg]^2 \Gamma_{h}^{\rm SM}(m_{\tilde{G}_{0^{++}}})~,
\end{equation}
where $\Gamma_{h}^{\rm SM}$ is the SM higgs decay width when $m_h=m_{\tilde{G}_{0^{++}}}$. The evaluated benchmark glueball lifetime at $m_0\gg 2m_b$ gives
\begin{equation}
c\tau_{\widetilde{G}_{0^{++}}\to {\rm SM}}\approx 0.28\left(\frac{f}{1~\text{TeV}}\right)^{4}\left(\frac{5~\text{GeV}}{\Lambda_d}\right)^{7}~{\rm cm}~,
\end{equation}
which is also consistent with~\cite{Cheng:2015buv}.
The expression is derived by assuming the existence of another heavy dark quark, like the twin-top, which carries a dark yukawa coupling with the same size of the SM top yukawa coupling. The heavy quark generates a dark gluon-fusion coupling between dark gluons and the dark Higgs, and the scalar glueball decays through the gluon-fusion coupling and the Higgs mixing. An example of the $\widetilde{G}_{0^{++}}$ lifetime bound is shown in Fig.~\ref{fig:chi} (right). The observed $\Omega_{\rm DM}h^2$ value requires $c\tau_{\widetilde{G}_{0^{++}}}\lsim 1$ cm for the mass splittings we assume. These cosmological constraints set upper bounds on the mediation scale $f$ and lower bounds on the dark confinement scale $\Lambda_d$.

\section{Application to the long-lived particle searches}\label{sec:search}
Here we study the probability of having LLP signatures inside particle detectors based on the cosmological bounds obtained in the previous section. There are several ways to produce dark hadrons at the LHC, and each production mode gives different boost factor distribution to the LLPs. Since the dark hadron boost determines the translation between the lifetime constraint and dark hadron's decay length in the lab-frame, we have to study the decay probability under different assumptions of hadron production mechanisms. 

To show the importance of the cosmological upper bounds on dark hadron lifetimes, we focus on collider searches using two smaller volume detectors
\begin{itemize}
\item LHCb VELO: consider both the pre- and post-module searches that require the LLPs decay within $0.1$-$22$~mm distance in the transverse direction with rapidity range $\eta\in [2,5]$~\cite{Ilten:2016tkc}.
\item ATLAS/CMS inner detector: we take a simplified assumption by requiring LLPs to decay within $1$-$30$~cm in the transverse direction with rapidity range $|\eta|<3$.
\end{itemize}
There are good motivations to use these tracker detectors for the LLP search. With a low $p_T$ triggering requirement and excellent particle identification ability, the LHCb detector is good at looking for light and soft LLPs~\cite{Pierce:2017taw}. The tracking information from the VELO detector is essential to reconstruct the decay location and veto hadronic backgrounds, and this is why most LHCb LLP searches have their best sensitivity for a $\mathcal{O}(1)$ cm scale decay length~\cite{Aaij:2017mic,Aaij:2016xmb,Aaij:2016isa}. Having inner detector information in ATLAS/CMS searches helps to reconstruct displaced decay signals~\cite{Sirunyan:2018pwn,Sirunyan:2018vlw,Aaboud:2017iio,Aaboud:2018jbr,Aaboud:2018iil,Sirunyan:2018njd}. The number of charged tracks from a hadronic LLP decay is also useful to distinguish the signal from QCD backgrounds. For the displaced muon search that is useful for probing LLPs in the photon portal scenario, the tracking information is required to identify muons and measure muon energy.

We consider three types of dark hadron productions. For the photon portal scenario, we consider dark hadron produced from the $Z$ boson decay ($Z\to\omega_d\,\eta_d$), and the direct production and decay of dark photon ($Z_d\to\omega_d\,\eta_d$). For the Higgs portal scenario, we consider dark hadron production from a Higgs decay ($h\to 2\chi_d$ or $2\widetilde{G}_{0^{++}}$). When calculating the decay length, we include the boost distribution of dark hadrons obtained from MadGraph5~\cite{Alwall:2011uj} simulations. In these studies we assume the energy cuts required in real searches do not change the boost distribution significantly. This is a valid assumption when considering existing searches from Higgs decays at ATLAS/CMS~\cite{Aaboud:2018aqj,Aaboud:2018jbr,Aaboud:2018iil} and the LLP searches at the LHCb~\cite{Ilten:2016tkc,Pierce:2017taw,Aaij:2016isa}, which do not require high $p_T$ final states in the searches. We also do not consider scenarios that allow long-lived particles to be trapped inside the detector and decay after a long time~\cite{Sirunyan:2017sbs} and assume dark hadrons simply fly away from the production point.

\subsection{Vector meson decay in the photon mixing scenario} 
The kinetic mixing in Eq.~(\ref{eq:kinmix}) introduces a coupling between the SM $Z$ and dark quarks, which makes $Z_d^{\mu}\to Z_d^{\mu}+\theta_Z\,Z^{\mu}$ after redefining gauge bosons by shifting away the kinetic mixing. The mixing angle is written as~\cite{Blinov:2017dtk}
\begin{eqnarray}
\theta_Z \simeq \frac{\epsilon \tan\theta_W m_Z^2}{m_Z^2 -m_{Z_d}^2}\,\,,
\end{eqnarray}
and $\theta_W$ is the weak mixing angle.
\begin{figure}
\begin{center}
\includegraphics[scale=0.36]{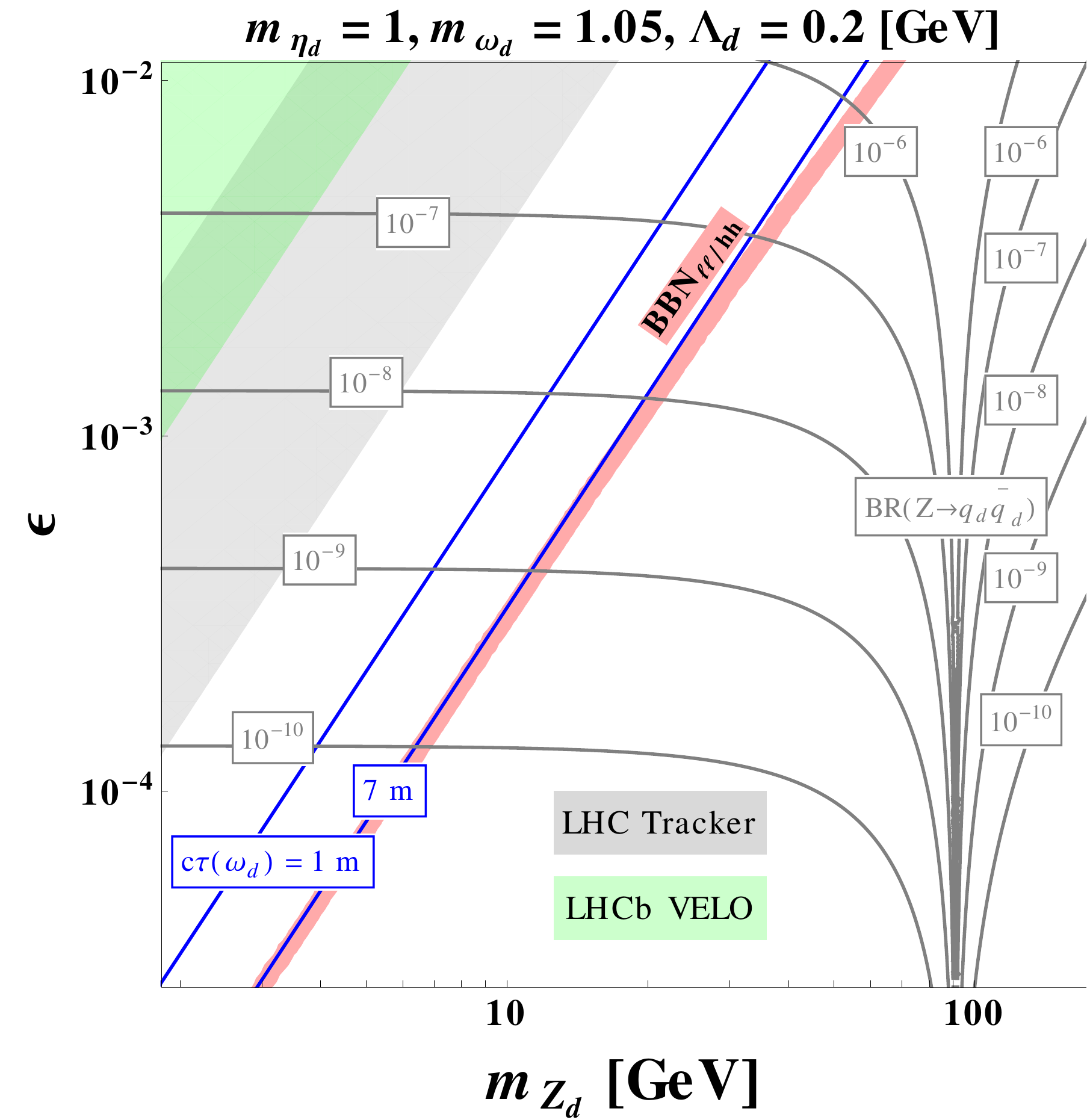}\qquad
\includegraphics[scale=0.37]{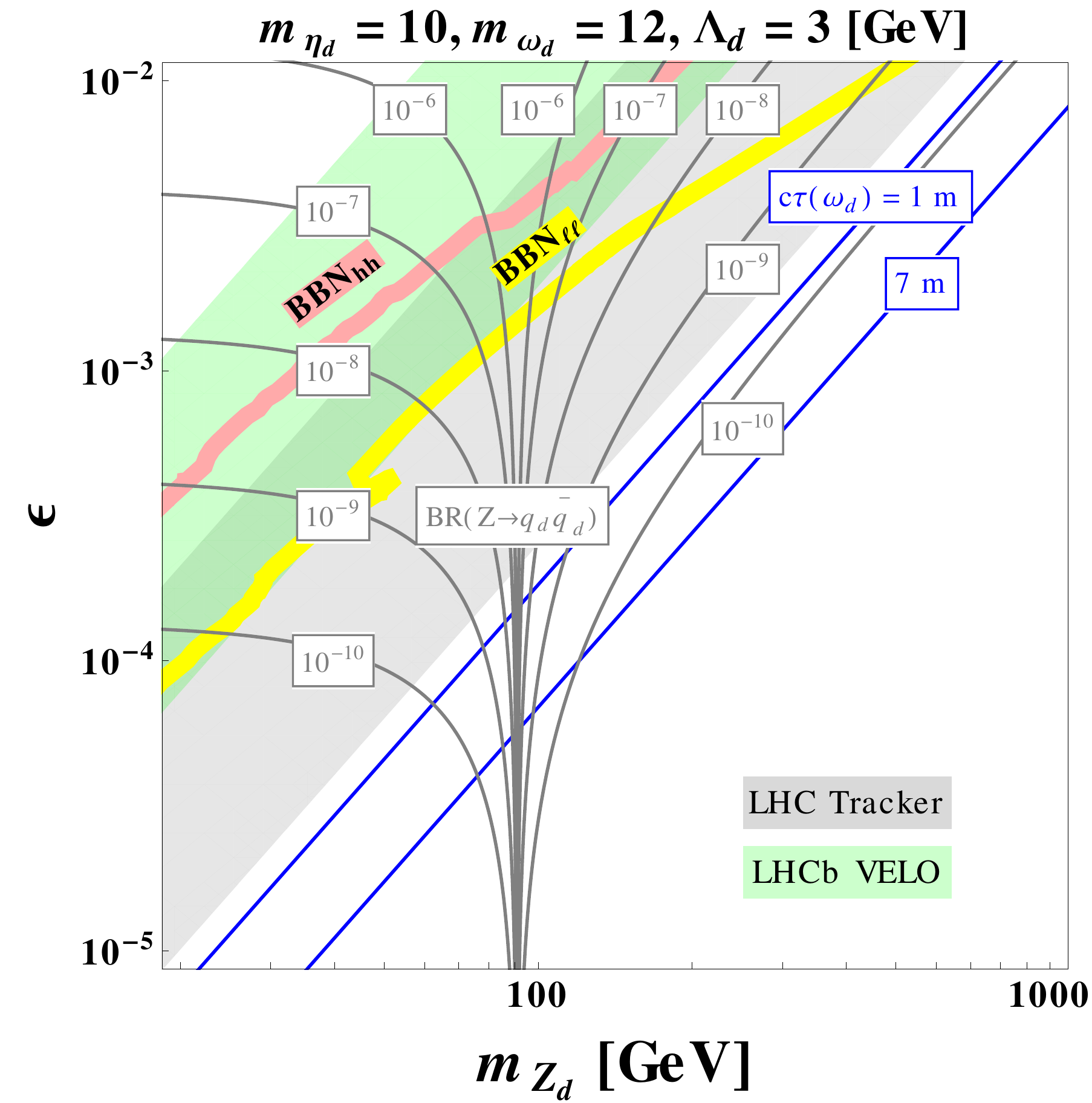}
\end{center}
\caption{Comparisons between detector size and BBN bounds on the $\omega_d$ lifetime. We consider a $14$ TeV search of exotic $Z$ decay into $\tw+$MET, and $\omega_d$ has a displaced decay into $\mu\mu$. The BBN bounds (red and yellow curves) favor the upper left region of the parameter space, which allows quick $\omega_d$ decays to suppress $\eta_d$ density. The green (gray) shaded region corresponds to the parameter space of having $>10\%$ probability for $\omega_d$ to decay inside the LHCb VELO (ATLAS/CMS inner detector), with the approximated detector geometries assumed in the beginning of Sec.~\ref{sec:search}. The decay probability takes into account the boost factor and angular distribution of $\omega_d$ obtained from MadGraph5 simulations. Two blue curves $c\tau(\tw)=1$ and $7$~m roughly correspond to the distance from the primary vertex to the ECAL and muon chamber in ATLAS/CMS. We show contours of BR$(Z\to q_d\bar{q}_d)$ assuming dark photon coupling $\alpha_d=\alpha$.}
\label{fig:LLP_Geo1}
\end{figure}

The mixing turns on the decay $Z\to q_d\,\bar{q}_d$, and we show the decay branching ratio in black curves of Fig.~\ref{fig:LLP_Geo1} by assuming dark quarks coupling to dark photon with $\alpha_d=\alpha$. Once dark quarks are produced, they form dark hadron final states including $Z\to\omega_d\,\eta_d$ through the dark hadronization process. Since $\eta_d$ is much longer-lived, the event contains missing energy plus a displaced decay of $\omega_d$ into SM fermions. The kinetic mixing gives $\omega_d$ an $\mathcal{O}(0.1)$ branching ratio to decay into muons, and the displaced muon signals that generate tracks inside the LHCb VELO or ATLAS/CMS tracker give clean LLP signatures.

In Fig.~\ref{fig:LLP_Geo1}, we show two examples of the dark meson spectrum and mark the $Z_d$ mass and coupling that make $\omega_d$ decay inside the VELO or tracker detectors. The green (gray) shaded region correspond to having $\geq10\%$ probability for $\omega_d$ to decay in the pre- or post-module LHCb search region (tracker search region). These decay regions should be compared to the parameter space above the pink (yellow) curve that satisfies the BBN constraints if $\eta_d$ decays hadronically (leptonically). For the parameters used in the left panel, where the dark hadron mass gap is small compared to their masses, the cosmological arguments are less restrictive, and the decay length can be comparable to the size of muon spectrometer in ATLAS/CMS ($\sim4$-$10$ m).  For the benchmark shown in the right panel, the BBN bound requires $c\tau(\tw) \lesssim 1$~cm, and most of the decays show up in the VELO and other inner detectors.

Since $\omega_d$ has a good chance to decay inside the LHCb VELO, we can estimate the size of BR$(Z\to q_d\bar{q}_d)$ that can be probed in the near future. LHCb is going to produce $\approx 8\times 10^8$ $Z$'s with $15$ fb$^{-1}$ of data~\cite{Zakharchuk:2308473}. After taking into account the $10\%$ decay probability for the VELO detector (green region) and $\sim 50\%$ reconstruction efficiency of a displaced muon pair~\cite{Ilten:2016tkc} times an additional $15\%$ branching ratio of $\omega_d\to\mu\mu$, we can cover the $Z\to q_d\bar{q}_d$ branching ratio down to $\sim 10^{-6}\,(2\sigma)$ from the LHCb search. The estimate assumes this $Z$ decay to always produce an $\omega_d$ in the final state, and there are $\sim 10$ backgrounds per $\omega_d$ mass bin in the search. The choice of the number of background is motivated by the estimate in Ref.~\cite{Ilten:2016tkc} that discusses the displaced muon search from dark photon decays. This branching ratio region does exist in the upper left corner of the plots in Fig.~\ref{fig:LLP_Geo1}, which satisfies the BBN constraint with a GeV scale $\eta_d$-$\omega_d$ splitting and have $\omega_d\to\mu\mu$ inside the VELO search region. For a given $Z_d$ mass, the BBN constraint also sets a lower bound on $\epsilon$ that is complementary to the collider bound for a larger mixing.

For the search at ATLAS/CMS, we can estimate the future sensitivity of BR$(Z\to q_d\bar{q}_d)$ based on an existing ATLAS search~\cite{Aaboud:2018jbr}. The search has a collimated di-muon trigger ($\Delta R_{\mu\mu}<0.5$) with relatively low muon energy requirement ($p_T>15$ and $20$ GeV) and therefore has a good chance of picking up the $\omega_d$ signal. The efficiency of selecting a displaced di-muon vertex is much higher when the transverse impact parameter of the muons is between $10$ to $50$ cm. This overlaps with the gray region ($1$-$30$ cm) in Fig.~\ref{fig:LLP_Geo1}. Using the result in Ref.~\cite{Aaboud:2018jbr} for the $h\to Z_d\,Z_d$ search that reconstructs one of the displaced $Z_d\to\mu^+\mu^-$ decays, we rescale their best cross section bound for $20$ GeV $Z_d$ with $c\tau_{Z_d}=10$ cm by the $Z$ production rate, and the expected BR$(Z\to q_d\bar{q}_d)$ bound is $\sim 10^{-7}$ ($2\sigma$) with $300$ fb$^{-1}$ of data. The $10$ cm decay length is less preferred by the BBN constraint in Fig.~\ref{fig:LLP_Geo1} for a GeV scale meson splitting (like the gray region in Fig.~\ref{fig:LLP_Geo1} right), but the signal can exist for a more degenerate meson spectrum (gray region in Fig.~\ref{fig:LLP_Geo1} left).

\begin{figure}
\begin{center}
\includegraphics[scale=0.36]{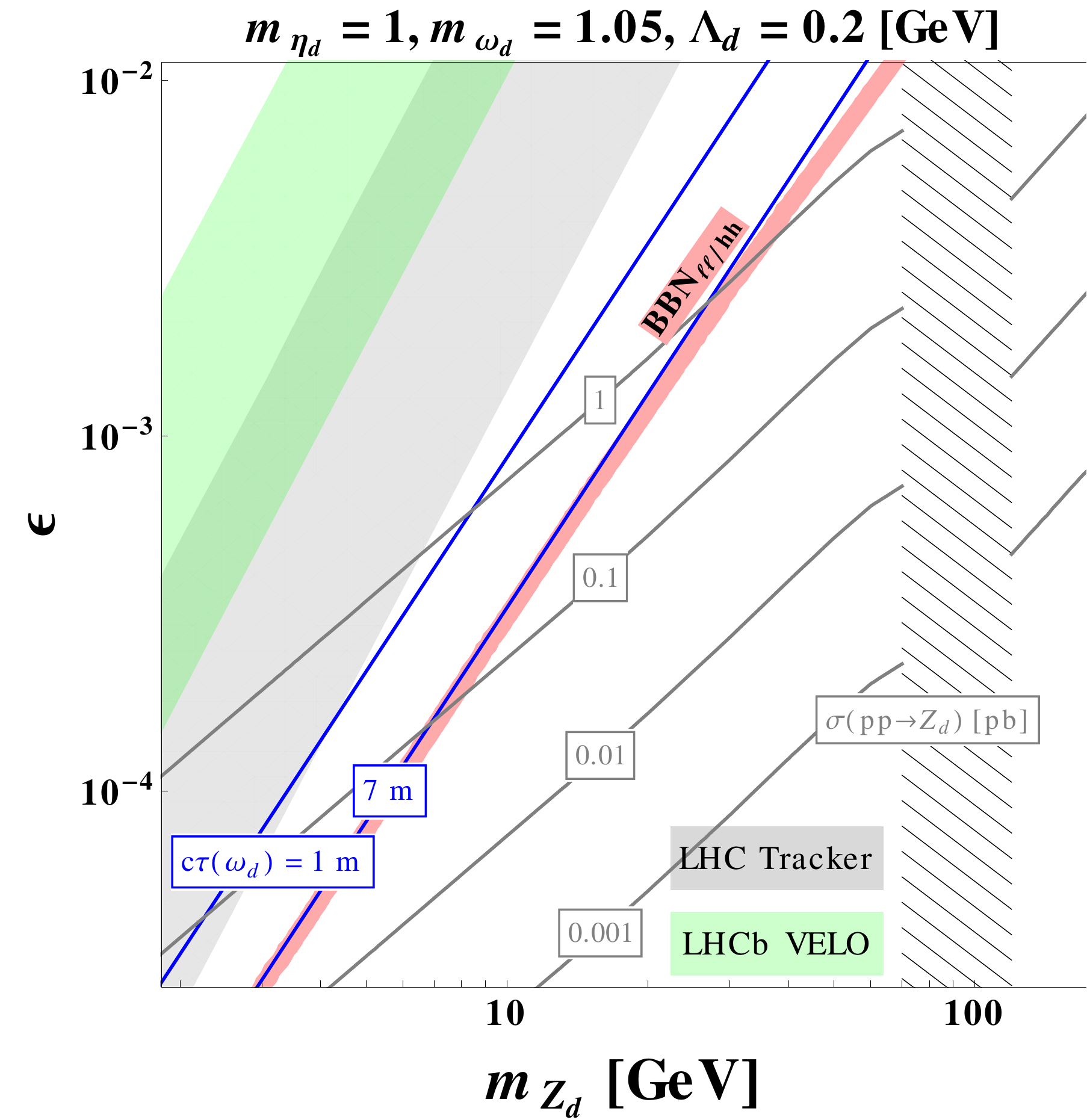}\qquad
\includegraphics[scale=0.37]{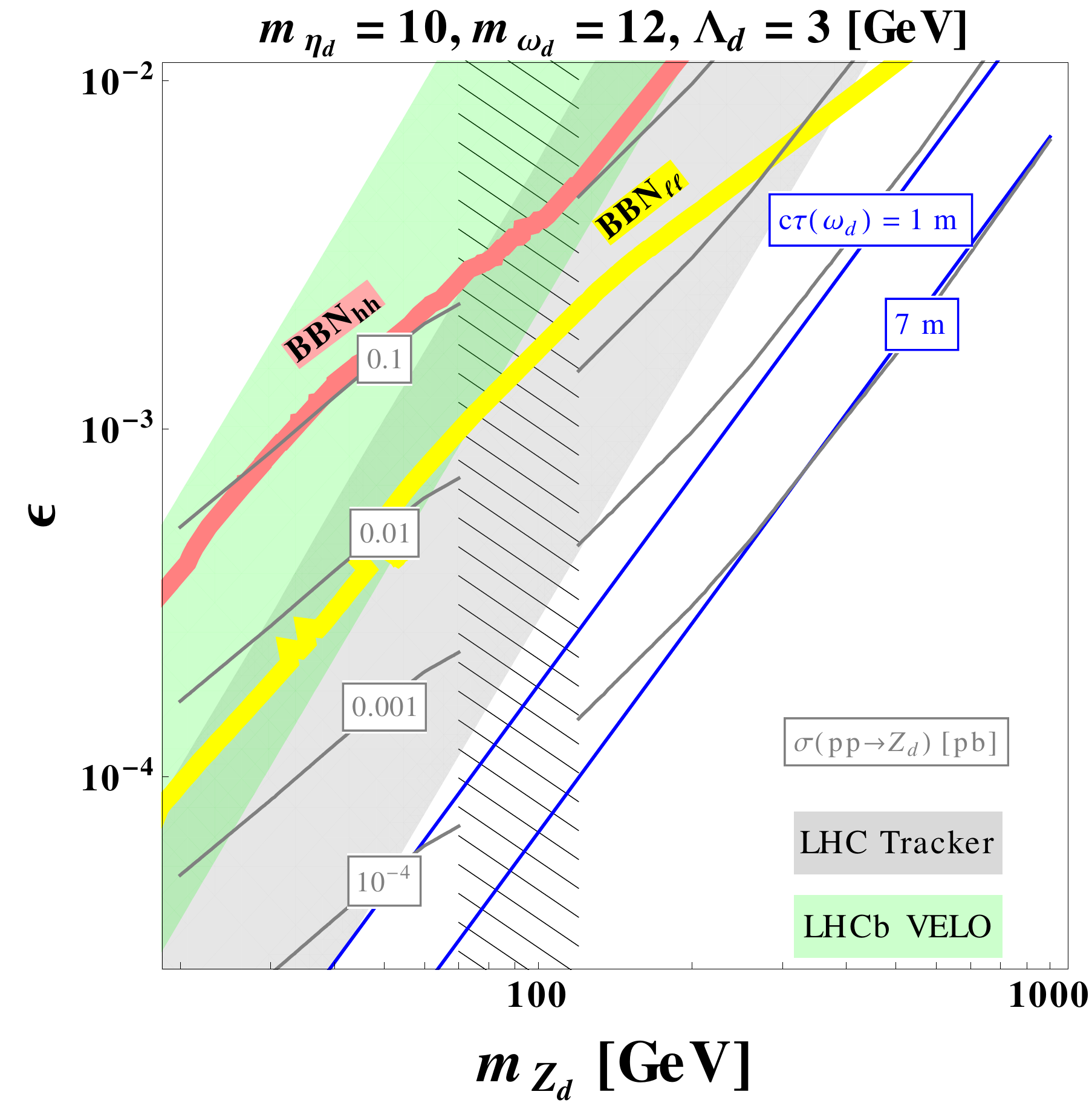}
\end{center}
\caption{Similar to~Fig.\ref{fig:LLP_Geo1} but with a different $\omega_d$ production process, $pp\to Z_d\to \omega_d+$MET. The BBN bounds favor the parameter space in the upper left corner, which gives the right decay length and larger $Z_d$ production for the LHCb and ATLAS/CMS searches using tracker detectors. We do not show the dark meson production near the $Z$ peak (hatched region).}
\label{fig:LLP_Geo2}
\end{figure}

Besides the $Z$ decay, we can also produce $\omega_d$'s through the direct $Z_d$ production from SM quarks. In Fig.~\ref{fig:LLP_Geo2}, we show the production cross section of $Z_d$ (black curves) and the green-/gray-filled regions with the same definition in Fig.~\ref{fig:LLP_Geo1}. Since $Z_d$ dominantly decays into dark hadrons that generate missing energy or LLP signatures, most of the dark photon constraints obtained by looking at prompt and visible $Z_d$ decays do not apply~\cite{Curtin:2014cca,Aaboud:2018fvk,Sirunyan:2018mgs}. The search of displaced $Z_d$ decay, such as~\cite{ATLAS-CONF-2016-042,Aaij:2017rft}, also have not covered the $\omega_d$ mass and lifetime we consider. However, since the BBN constraints require $\gsim100$ fb production of long-lived $\omega_d$'s that decay inside the VELO and the inner detectors, future improvements on the displaced $Z_d$ search may cover the parameter space that is complementary to the BBN bounds\footnote{See, e.g.,~\cite{Pierce:2017taw,Tsai:2018vjv} for the projection of future LHCb/ATLAS/CMS bounds on heavy resonances decaying into LLPs that decay leptonically.}.

\subsection{Scalar hadron decay in the Higgs portal scenario}
For the Higgs portal scenario, we focus on LLP signals from the Higgs boson decay into two scalar hadrons, either $2\chi_d$ or $2\widetilde{G}_{0^{++}}$. Although it is possible to get final states with higher multiplicity from the hidden QCD process, we first focus on the simplest case by assuming $1\to 2$ decays. We comment on the dark shower signals that contain multiple final state particles in the next sub-section.  The decay of $\mathcal{O}(1{\rm-}10)$ GeV scalar hadrons is mainly into SM quarks through the Higgs portal coupling, and the collider signatures can be displaced jets or hadrons. 

\begin{figure}
\centering
\includegraphics[scale=0.37]{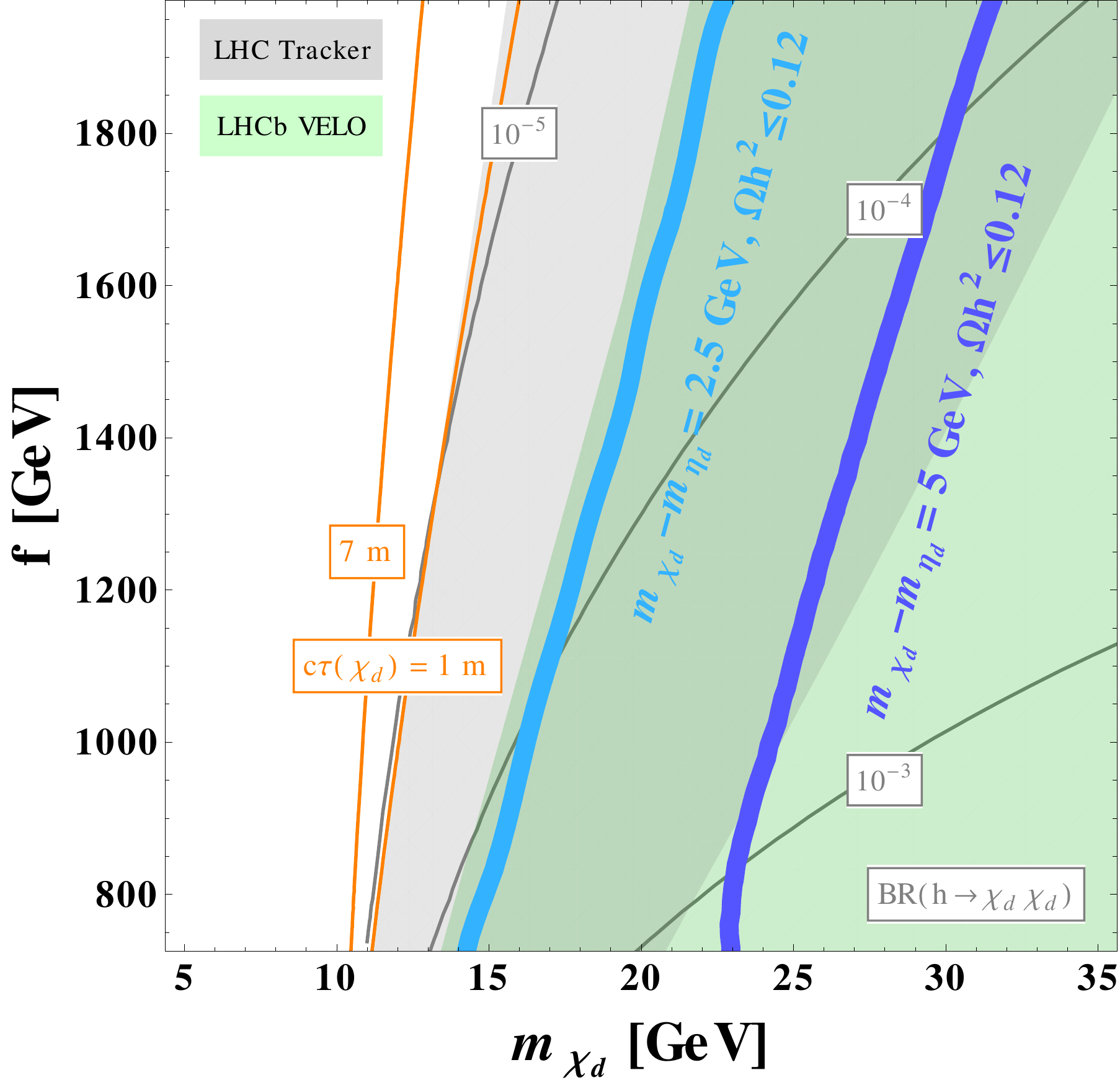}\qquad
\includegraphics[scale=0.37]{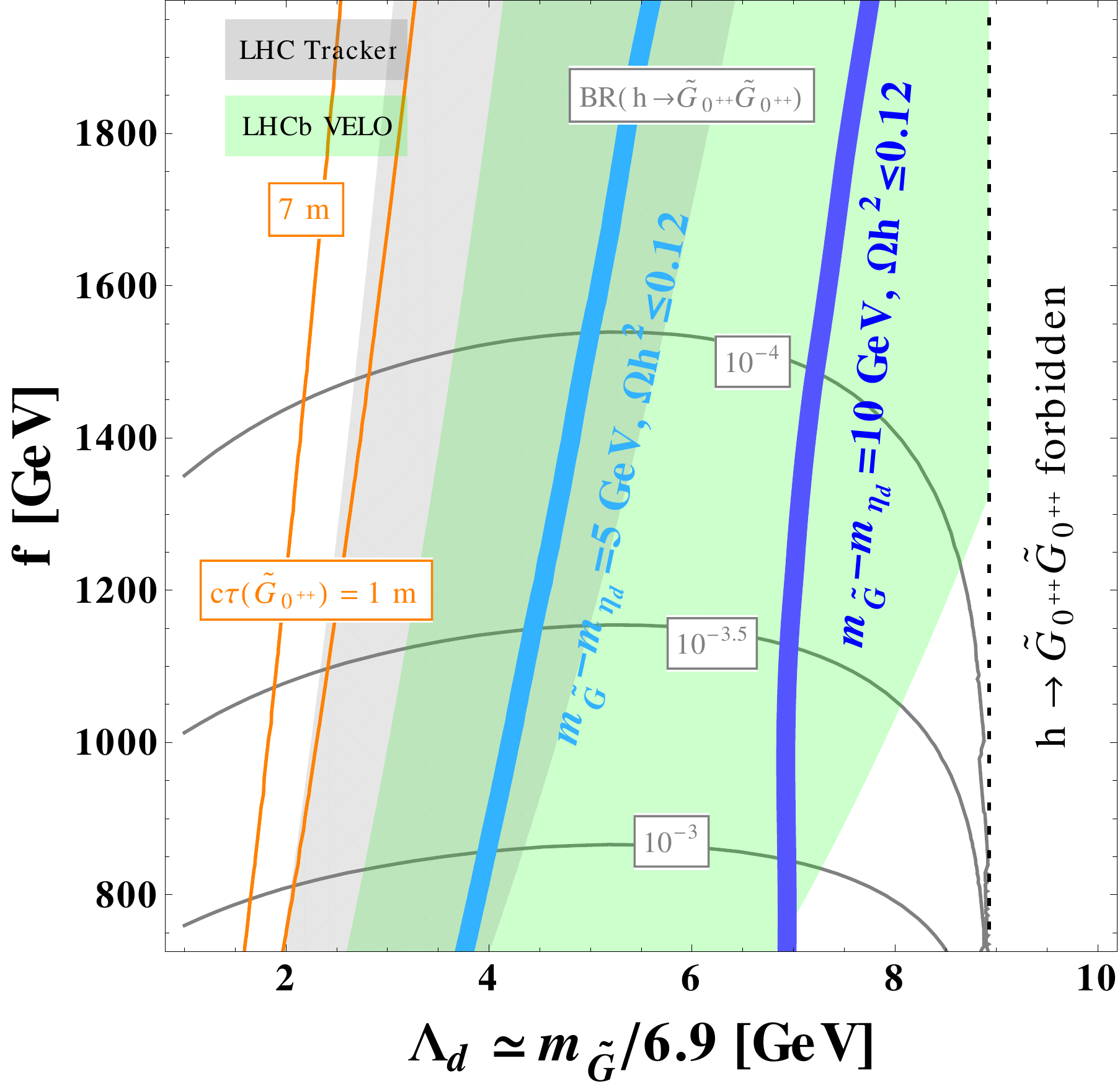}
\caption{Comparisons between detector size and the $\Omega_{\rm DM}h^2$ bound on scalar hadron lifetimes. We consider a $14$ TeV search of exotic Higgs decay into $2\chi_d$ (left) or $2\widetilde{G}_{0^{++}}$ (right), and the scalar hadrons decays into SM quarks through a Higgs portal coupling. The BBN bounds (light- and dark-blue curves) favor the parameter space on the right, which allows quick enough scalar hadron decays to suppress $\eta_d$ density. The green (gray) shaded region corresponds to the parameter space of having $>10\%$ probability for the scalar hadrons to decay inside the LHCb VELO (ATLAS/CMS inner detector), with the approximated detector geometries assumed in the beginning of Sec.~\ref{sec:search}. The decay probability takes into account the boost factor and angular distribution of $\omega_d$ obtained from MadGraph5 simulations. Two orange curves $c\tau(\tw)=1$ and $7$~m roughly correspond to the distance from the primary vertex to the ECAL and muon chamber in ATLAS/CMS. We show contours of BR$(h\to 2\chi_d)$ by assuming a $10\%$ chance of producing $2\chi_d$ from $q_d\bar{q}_d$. The BR$(h\to 2\widetilde{G}_{0^{++}})$ is estimated by assuming all the Higgs decay into dark gluons form $2\widetilde{G}_{0^{++}}$, and the decay into $q_d\bar{q}_d$ only gives missing energy.
}\label{fig:LLPglueball}
\end{figure}

In Fig.~\ref{fig:LLPglueball}, we show the parameter space for having more than $10\%$ decay probability for the scalar hadrons to show up in the pre-/post-module LHCb search (green) or the inner detector search from ATLAS/CMS (gray) defined earlier in this section. In these plots, the larger confinement scale or heavier $m_{\chi_d}$ in the glueball or meson case corresponds to faster scalar hadron decays. Constraints from requiring $\Omega_{\eta_d}<\Omega_{\rm DM}$ prefers the parameter space in right side of the light-/dark-blue curves, and the location of DM density bound depends on the mass splitting between dark hadrons. When plotting the bounds, we keep the micro-physics parameters that determine the hadron mass splittings implicit, and the cosmological bound is insensitive to those details. In the $\chi_d$ plot, we take the mass and lifetime assumptions in Sec.~\ref{sec:Higgsportal} when calculating the bounds. In the $\widetilde{G}_{0^{++}}$ plot, each of the parameter point has a corresponding range of $y_d$ that gives the mass hierarchy $m_{\eta_d}<m_{\widetilde{G}_{0^{++}}}<m_{\chi_d}$ so the glueball plays a major role in the annihilation/decay process. Once being produced at colliders, the scalar hadrons prefer to decay inside the LHCb VELO and ATLAS/CMS inner detectors according to the cosmological bounds from the light-/dark-blue curves.

To get an idea of the experimental coverage of the parameter space, we estimate the size of Higgs decay branching ratio into scalar hadrons and compare the result to bounds from LLP searches. However, since the result depends on details of the dark quarks, we need to make some assumptions of hidden particles when showing the branching ratio. In the $\chi_d$ scenario, we assume the dominant exotic decay channel of Higgs is into dark quarks. We follow the same $m_{\chi_d,q_d}$ assumption in Eq.~(\ref{eq:mesonm}) and choose the dark confinement scale $\Lambda_d=5$ GeV in the calculation. The Higgs mixing introduces the $h\to q_d\bar{q}_d$ decay, and the decay branching ratio is estimated as
\begin{equation}
\Gamma_{h\to 2\chi_d} \approx A \bigg(\frac{m_{q_d}}{m_b} \frac{v^2}{f^2} \bigg)^2 \Gamma_{h\to bb}(\text{SM}),
\end{equation} 
where $A$ can be an $\mathcal{O}(0.1)$ factor for forming $\chi_d$'s among all the dark hadrons. In Fig.~\ref{fig:LLPglueball} (left) we show the branching ratios by taking $A=0.1$.

In the glueball scenario, we assume SM Higgs has exotic decays dominantly into a pair of $q_d$'s or a pair of dark gluons. We assume the decay into dark gluons always gives $2\,\widetilde{G}_{0^{++}}$, and the decay into dark quarks always form $\eta_d$'s that show up as missing energy. The partial widths into scalar glueballs can be estimated as~\cite{Craig:2015pha}
\begin{equation}
\Gamma_{h\to 2\widetilde{G}} \approx \bigg(\frac{\hat{\alpha}_s}{\alpha_s} \frac{v^2}{f^2} \bigg)^2 \Gamma_{h\to gg}(\text{SM}).
\end{equation}
$\hat{\alpha_s}$ and $\alpha_s$ are the gluon couplings in the dark and SM sectors at the Higgs mass scale, and we follow the same assumption of glueball coupling in Sec.~\ref{sec:Higgsportal}. 
We calculate BR$(h\to2\widetilde{G}_{0^{++}})$ using the partial width and show the result in Fig.~\ref{fig:LLPglueball} (right).

The existing CMS search on displaced jets signatures~\cite{Sirunyan:2018vlw} using $36$ fb$^{-1}$ of data has its best $2\sigma$ constraint on $50$ GeV LLP production down to $\approx20$ fb when $c\tau\approx 1$ cm. When applying the result to LLPs from Higgs decay, the search is already sensitive to Higgs BR$\sim10^{-3}$. This covers part of the parameter space in Fig.~\ref{fig:LLPglueball} that satisfies the cosmological bound for a GeV-scale mass splitting, and the bound will be improved when having more data. However, Ref.~\cite{Sirunyan:2018vlw} does not show results with the lighter LLP mass we consider, and the bound can be weaker due to the larger multi-jet backgrounds for lower energy signals. According to the projection in~\cite{Bediaga:2018lhg} assuming $300$ fb$^{-1}$ of data, the LHCb search may probe the Higgs branching ratio down to $10^{-3}$ level if the LLP has $\approx 50$ GeV mass and decay length $c\tau\approx 1$ mm. This covers the parameter space satisfying the cosmological bound with $>5$ GeV mass splitting.

\subsection{Dark shower signals}
Here we comment on the dark shower process that produces LLPs with high multiplicity. In a confining hidden valley model, the showering and hadronization process from dark QCD is likely to generate more than two dark hadrons, and many phenomenological studies have been proposed to look for these signals with multiple displaced vertices and softer final state particles~\cite{Schwaller:2015gea,Cohen:2015toa,Pierce:2017taw,Renner:2018fhh}. Similar to the SM QCD, a reasonable dark showering process should generate an energy distribution of dark hadrons that peaks at the infrared. This means no matter what kind of resonance decay used for generating dark showers, there is always a significant fraction of dark hadrons having small  boost. Therefore, the proper lifetime bounds we derived using BBN and $\Omega_{\rm DM}h^2$ can be easily applied to dark shower signals. Moreover, the requirement of $c\tau\lsim 1$ m from the cosmological bounds means a dark shower event is likely to contain multiple displaced decays inside the LHCb and ATLAS/CMS searches. This motivates the search of using more displaced vertices in one event to veto the background.

\section{Conclusion}
\label{sec:conclusion}
In this paper, we study the thermal history of $\mathcal{O}(1-10)$ GeV scale dark hadrons in a confining hidden valley model, in which the lightest dark meson has a slow decay that can violate the BBN or DM density constraints. As we show, in order to satisfy the cosmological bounds, one of the heavier dark hadrons need to decay into SM particles within $\sim10^{-8}$ sec lifetime. This cosmological constraint does not only motivate the LLP search using meter-scale size detectors, but also suggests lower bounds on the dark hadron production that are complementary to collider constraints.

We explain the idea using two CHV scenarios: one with a photon portal coupling, and one with a Higgs portal coupling between the SM and dark particles. In the photon portal scenario, we show that the BBN constraint suggests the dark vector meson to have a good chance to be seen at the LHCb and ATLAS/CMS from the exotic $Z$-decay, and the produced vector meson $\omega_d$ mainly decays inside the LHCb VELO and the ATLAS/CMS inner detectors. In the Higgs portal scenario, the lightest meson $\eta_d$ can obtain a DM density from the conversion/decay process, and the observed $\Omega_{\rm DM}$ requires the lightest scalar hadron ($\chi_d$ or $\widetilde{G}_{0^{++}}$) to have $\lsim 10$ cm scale proper decay length for GeV-scale hadrons. Similar to the photon portal scenario, the DM density constraint also suggests a sizable scalar hadron production from the Higgs decay, and the produced scalar hadrons are likely to decay inside the VELO and the inner detectors. The same study can be applied to LLP searches at proposed future detectors~\cite{Curtin:2018mvb,Feng:2017uoz,Alekhin:2015byh,Gligorov:2017nwh,Liu:2018wte}, and their different coverage of LLP's decay length is complementary to the cosmological bounds with different hadron mass splittings.

Although the lifetime bounds we show only assume $\mathcal{O}(0.1-1)$ GeV mass splittings between dark hadrons, the same numerical analysis can be applied to different CHV scenarios even outside of this mass range. For some well-motivated scenarios such the Twin Higgs model, we have better defined hidden sector parameters and can estimate the twin meson masses and lifetimes for studying the cosmological constraint~\cite{Cheng:2015buv}. Besides numerically solving the Boltzmann equations, the analytical approximation using Eqs.~(\ref{eq:Tdec}, \ref{eqn:energy_exchange_1}, \ref{eq:Ylestimate}) also gives an idea of the lightest meson abundance before its later decay. The result can be translated into the lifetime constraint after comparing to the BBN and $\Omega_{\rm DM}h^2$ bounds.

Another assumption we make in the study is that $\eta_d$ cannot simply decay into dark radiation to avoid cosmological bounds. This assumption is easily realized in the Higgs portal scenario since the CP odd $\eta_d$ cannot decay in SM particles through the Higgs coupling. In the photon portal scenario, however, $\eta_d$ may decay quickly into dark photons if the process is kinematically allowed. For example, the cosmological bound is much weaker if the dark photon mass is between 10 MeV$\lsim m_{Z_d}<\frac{1}{2}m_{\eta_d}$, so $\eta_d\to 2Z_d(e^+e^-)$ can easily happen before the BBN. For an even lighter $Z_d$, since it cannot decay into electrons directly, we may still obtain useful bounds on the hadron lifetime if the much slower decay $Z_d\to 3\gamma$ injects energy to the thermal bath and violates the BBN and CMB bounds. We leave the possibility for future study. If the dark photon is massless, $\eta_d$ can quickly decay into dark photons to avoid the BBN bounds as long as the dark sector is colder than the SM sector and produces $\Delta N_{eff}\lsim0.4$~\cite{Aghanim:2018eyx}. However, the near future CMB S-4 experiments can probe $\Delta N_{eff}\gsim0.02$~\cite{Abazajian:2013oma}, and the null result will rule out this massless dark photon scenario unless a severe asymmetric reheating between the SM and dark sector happens after the two sectors decouple~\cite{Berezhiani:1995am,Chacko:2016hvu,Craig:2016lyx}. 

The general possibility of having a hidden sector that contains non-trivial forces and particle content has drawn considerable interest recently, due to its potential in providing solutions to physics puzzles and generating exotic collider and cosmological signatures. Our work provides an example of combining the collider and cosmological data, which usually correspond to physics in very different energy and time scales, to fully explore the parameter space of CHV sectors based on their unique thermal histories.

\section*{Acknowledgements}
We thank Hsin-Chia Cheng, Timothy Cohen, David Curtin, Thomas DeGrand, Johnathan Feng, Dan Hooper, Takeo Moroi, Michael Ramsey-Musolf, Ennio Salvioni, Haibo Yu, Keping Xie for useful discussions, and are especially grateful to Shmuel Nussinov for comments about the
manuscript. LL was supported by the General Research Fund (GRF) under Grant No 16312716, and by the US Department of Energy grant DE-SC-000999. YT was supported in part by the National Science Foundation under grant
PHY-1315155, and by the Maryland Center for Fundamental Physics. YT thanks the Aspen Center for Physics, supported by National
Science Foundation grant PHY-1607611, where part of this work was performed.

\bibliography{./Biblography}

\end{document}